\documentclass[amsmath,notitlepage,prd]{revtex4-1}
\usepackage{amssymb}
\usepackage{graphicx}
\usepackage{xcolor}
\usepackage{titlesec}
\usepackage{subfig}
\usepackage{multirow}
\usepackage{hhline}

\newcommand\lCompton{\lambda_{\mbox{\scriptsize Compton}}}
\newcommand\ldeBrogile{\lambda_{\mbox{\scriptsize de Brogile}}}
\newcommand\qqpair{Q\bar Q}
\newcommand\qqpairr{Q\bar Q[r]}
\newcommand\npme{\left\langle 0 \left| \mathcal O^{H}[r]\right| 0\right\rangle}
\newcommand\RR{\left|R'(0)\right|^2}
\newcommand\OS{\langle O_S \rangle}
\newcommand\OP{\langle O_P \rangle}

\newcommand\RRp[1]{\left| R_{ \chi_{ #1 } }' (0)\right|^2 }
\newcommand\OSp[1]{\langle O_S \rangle_{ \chi_{ #1 } }}
\newcommand\OPp[1]{\langle O_P \rangle_{ \chi_{ #1 } }}

\newcommand\OSRR{M^2\OS/\RR}
\newcommand\OPRR{\OP/\RR}

\newcommand\chirat[2]{\sigma(\chi_{#1})/\sigma(\chi_{#2})}

\begin{document}
\title{Production of $\chi_c$- and $\chi_b$-mesons in high energy hadronic collisions}

\author{A. K. Likhoded}
\email[E-mail: ]{\href{mailto:Anatolii.Likhoded@ihep.ru}{Anatolii.Likhoded@ihep.ru}}
\affiliation{Institute for High Energy Physics, 142281 Protvino, Moscow Region, 
Russia}
\affiliation{Moscow Institute of Physics and Technologe, Dolgoprudny, Russia}

\author{A. V. Luchinsky}
\email[E-mail: ]{\href{mailto:Alexey.Luchinsky@ihep.ru}{Alexey.Luchinsky@ihep.ru}}
\affiliation{Institute for High Energy Physics, 142281 Protvino, Moscow Region, 
Russia}
\affiliation{SSC RF ITEP of NRC ``Kurchatov Institute''}

\author{S. V. Poslavsky}
\email[E-mail: ]{\href{mailto:stvlpos@mail.ru}{stvlpos@mail.ru}}
\affiliation{Institute for High Energy Physics, 142281 Protvino, Moscow Region, 
Russia}
\affiliation{SSC RF ITEP of NRC ``Kurchatov Institute''}

\begin{abstract}
This paper is devoted to phenomenological study of $\chi_{c,b}$-mesons production in high energy hadronic collisions in the framework of NRQCD. We analyze all available experimental data on $\chi_c$-mesons production and extract non-perturbative NRQCD matrix elements from fitting the data. We show, that measured $p_T$-spectrum of $\chi_c$-mesons is mainly formed by color singlet components, while $\sigma(\chi_{c2})/\sigma(\chi_{c1})$ ratio depends strongly on color octet matrix elements; this ratio becomes a highly sensitive tool to study contribution of different terms from the NRQCD expansion. Obtained using NRQCD scaling rules predictions for $\chi_b$-mesons cross sections are also given.
\end{abstract}


\maketitle

\section{Introduction}
\label{sec_intro}
From the moment of discovery of the first heavy quarkonium state --- $J/\psi$-meson, this family of particles provided a very sensitive tool for both theoretical and experimental studies of QCD nature in its two distinct regimes: perturbative and non-perturbative. This remarkable feature of heavy quarkonium is the consequence of two distinguishing properties of these mesons. The first one is that the mass of heavy quarkonium $M$ is significantly greater then $\Lambda_{QCD}$. Since the production/annihilation of quark-antiquark pair $Q\bar Q$ occurs at short distances $\lCompton \sim 1/M$, this property allows one to consider this subprocess within the perturbative QCD. The second distinguishing feature is that the intrinsic quarks velocity $v$ is small compared to speed of light. On the one hand, this means that the hadronization of $\qqpair$ pair into observable occurs at large distances $\ldeBrogile \sim 1/(M v)$ and it is essentially non-perturbative process. On the other hand, this nonrelativistic nature 
of quarkonium bound state allows one to develop potential models or some effective theories for theoretical description of such processes. 

Historically, the first theoretical description of heavy quarkonium production was given by a {\itshape Color Singlet (CS) Model}  \cite{Kartvelishvili:1980uz,Baier:1983va}. In CS model it is assumed that heavy quarkonium bound state consists only of the $\qqpair$ pair in a colorless combination. The production amplitude can be factorized into two factors: amplitude describing production of $\qqpair$ pair within perturbative QCD and a probability amplitude which describes hadronization and is simply proportional to Schr{\H o}dinger's wave function of bound state (or its derivative) at the origin. The latter parameter can be calculated numerically within appropriate potential models, or can be extracted from e.g. experimental widths of mesons. Actually, the CS model takes into account only leading term in relative velocity of quarks. On the other hand, it is known that $v \sim \alpha_S(Mv)$, so the corrections bounded with finite value of $v$ are at least as important as corrections of order $\alpha_S(M) < \alpha_S(M v)$ in the perturbative part of total amplitude. The systematical accounting of $v$-corrections was later given by effective non-relativistic QCD (NRQCD) \cite{Bodwin:1994jh}. NRQCD assumes that in addition to pure $\qqpair$ colorless state quarkonium wave function contains states with $\qqpair$ pair in octet or singlet states accompanied by a dynamic gluons. Within the NRQCD the factorization formula can be written as follows:
\begin{equation}
\label{eq_nrqcd_main}
d \sigma\left(A+B \to H + X\right) = \sum_r d \sigma\left(A + B \to Q\bar Q[r] + X\right) \times \left\langle 0 \left| \mathcal O^{H}[r]\right| 0\right\rangle,
\end{equation}
where $H$ denotes a final particle, $r$ is a whole set of quantum numbers of $\qqpair$ pair ($n,S,L,J$ and color), and $\npme$ are the vacuum expectations of 4-fermionic operators arising in effective theory. These non-perturbative matrix elements absorbs physics of large distances responsible for $\qqpair$ pair hadronization, while the cross sections of $\qqpairr$ pair production describe physics of short-distances and are calculable within ordinary QCD perturbation theory. The main result of NRQCD is that   series \eqref{eq_nrqcd_main} can be organized in terms of relative velocity $v$, which is a small parameter. Thus, it is sufficient to consider only a fixed number of leading terms in series \eqref{eq_nrqcd_main}.

In the present paper we consider production of heavy quarkonium states $\chi_c$ and $\chi_b$ with high transverse momentum $p_T \gtrsim M$. This $p_T$ is achieved in short-distance part of reaction which is responsible for $\qqpair$ pair formation. It is well known, that in high energy hadronic reactions the dominant partonic mechanism of $\qqpair$ pair formation is gluonic fusion which leads to a reaction $gg\to \qqpair + g$ in the leading order in $\alpha_S(M)$. For $\chi_{c,b}$-states NRQCD predicts that both color singlet $\qqpair[^3P_J]$ and color octet $\qqpair [^3S^{(8)}_1]$ states contribute in the same order of $v$ in the total production rate. These two states have significantly different $p_T$-behavior and can be well separated by analyzing the available experimental data. As it was shown in our previous works \cite{Likhoded:2012hw, Likhoded:2013aya}, the available experimental data can not be explained well by considering only these two states, so contribution from higher octet states should also be considered. In the present paper we perform a rigorous fit of available experimental data including new LHC results and determine both central values and uncertanties for nonpeturbative model parameters. It will be shown that within  NRQCD and LO $\alpha_S(M)$ approximation the existing data on $\chi_c$ production can be separated into two groups that give two disjoint regions for non-perturbative parameters obtained using $\chi^2$-fit. We show, that measured $p_T$-spectrum of $\chi_c$-mesons is mostly formed by color singlet term, while higher octet states affect on the $p_T$-dependence of $\sigma(\chi_{c2})/\sigma(\chi_{c1})$. Then we use NRQCD scaling rules to obtain predictions for $\chi_b$-mesons cross sections. 

The rest of the paper is organized as follows. In the next section we briefly discuss theoretical framework used in our paper. In Sec.~\ref{sec_charm} we present the analysis of available experimental data for charmonium and determine contributions of color singlet and octet components into $\chi_{c1,2}$ production cross sections. Theoretical predictions for $\chi_{c0}$ production cross sections and $p_T$-dependence of the ratios $\sigma(\chi_{cJ_1})/\sigma(\chi_{cJ_2})$ are also given there. In Sec.~\ref{sec_bottom} we use NRQCD scaling rules to obtain same predictions for bottomonium. Brief analysis of our results is given in the conclusion.

\section{Theoretical background}
\label{sec_theory}
As it was discussed in the Introduction, the NRQCD factorization formula \eqref{eq_nrqcd_main} allows one to separate physics of short distances responsible for $\qqpairr$ pair production and physics of large distances responsible for hadronization. In the case of $pp$-scattering equation \eqref{eq_nrqcd_main} can be written as 
\begin{equation}
\label{eq_nrqcd_main_pp}
d \sigma\left(p+p \to H + X\right) = \sum_r \left( \int d x_1 \,d x_2 \,f_g(x_1; \mu^2) \,f_{g}(x_2;\mu^2) \,d \hat \sigma\left(g + g \to Q\bar Q[r] + g\right) \right) \times \left\langle 0 \left| \mathcal O^{H}[r]\right| 0\right\rangle,
\end{equation}
where $f_g(x; q)$ are gluon distribution functions of initial protons and $d \hat \sigma$ is the cross sections at  the partonic level. Using the NRQCD velocity scaling rules one can select the following relevant terms from the Fock structure of quarkonium bound state:
\begin{equation}
\label{eq_fock_structure}
\left| \chi_J \right \rangle = O(v^0) \left| \qqpair [^3P_J^{(1)}] \right \rangle + O(v^0)  \left| g \qqpair [^3S_1^{(8)}] \right \rangle  + O(v^2)  \left| g \qqpair [^1P_1^{(8)}] \right \rangle + O(v^2) \sum_{J'} \left| gg \qqpair [^3P_{J'}^{(8)}] \right \rangle + O(v^4),
\end{equation}
where superscript $(1,8)$ corresponds to the color of $\qqpair$ pair, and each big $O(v^n)$ before corresponding state indicates the relative contribution of this state in the total cross section \eqref{eq_nrqcd_main_pp} (i.e. relative powers of $v$ of corresponding matrix elements $\npme$). The first term in \eqref{eq_fock_structure} matches CS model. Second, color octet term, corresponds to the chromo-electric (E1) transition of $\qqpair$ pair into observable meson and contributes at the same order of $v$ (this is because CS term is in $P$-wave which brings additional power of $v$). The rest two terms, that contribute at the next order in $v$, correspond to chromo-magnetic (M1) and double chromo-electric (E1$\times$E1) transitions respectively. The phenomenological value of $v^2$ is about 0.3 and 0.1 for charmonium and bottomonium respectively, so these higher octet terms should give a noticeable corrections to leading order NRQCD terms.

\begin{figure}
\centering
\subfloat[]{\includegraphics[width=0.2\textwidth]{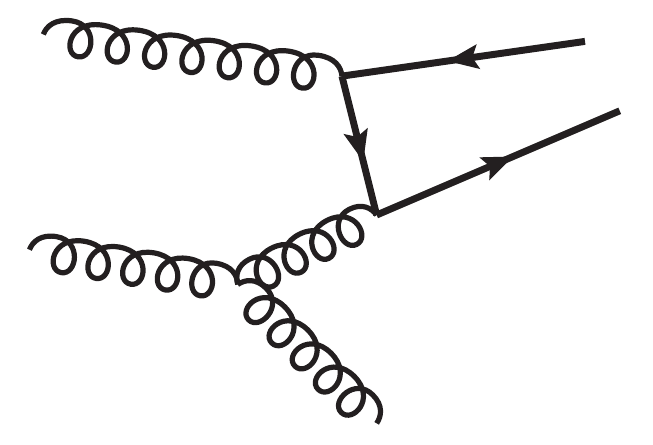}} 
\subfloat[]{\includegraphics[width=0.2\textwidth]{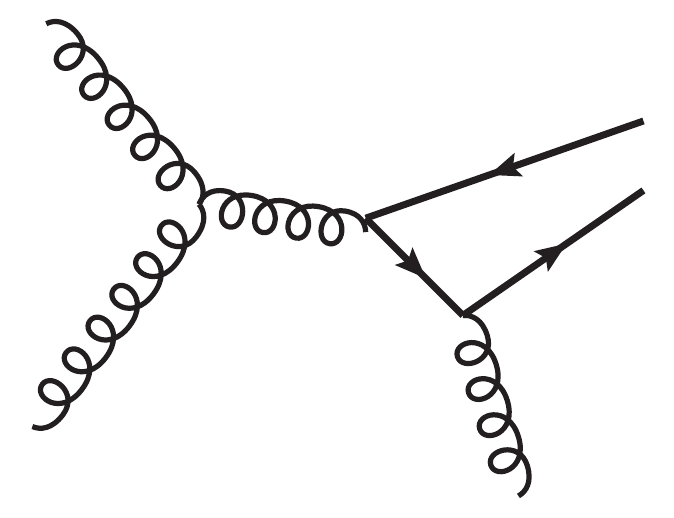}}\qquad
\subfloat[]{\includegraphics[width=0.2\textwidth]{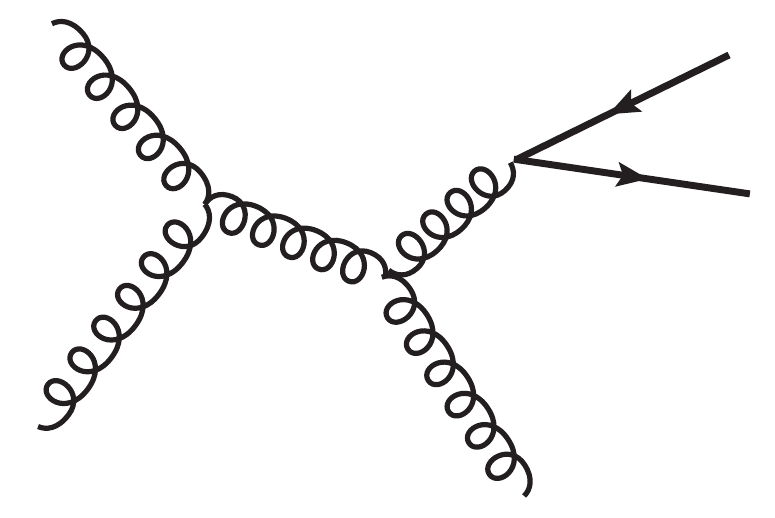}} 
\subfloat[]{\includegraphics[width=0.2\textwidth]{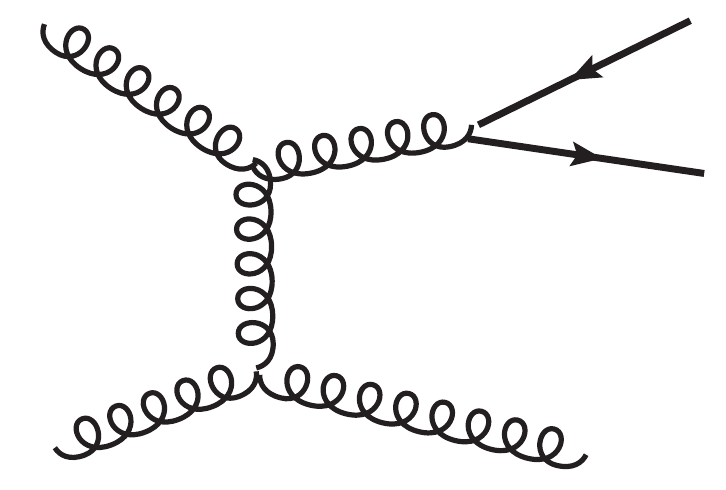}}\\
\subfloat[]{\includegraphics[width=0.2\textwidth]{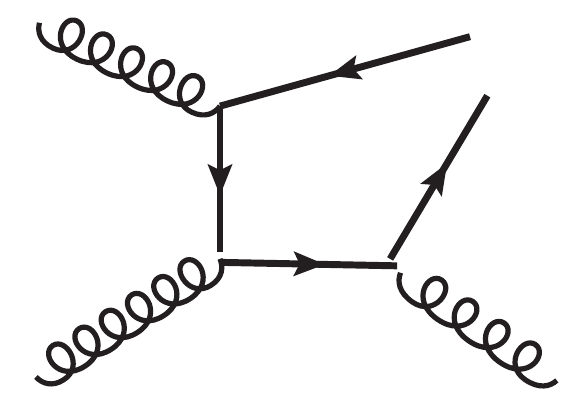}}\hspace{4cm} 
\subfloat[]{\includegraphics[width=0.2\textwidth]{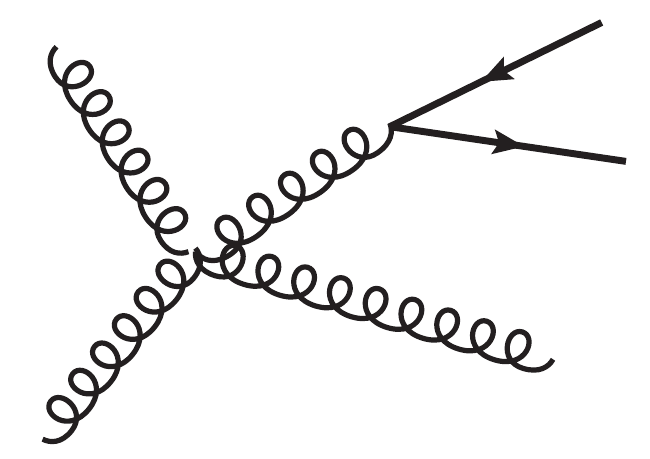}} 
\caption{Typical Feynman diagrams for the processes $g + g \to \qqpairr + g$. Diagrams (a), (b) and (e) corresponds to both color singlet and color octet production, while diagrams (c), (d) and (f) only for color octet production of $^3S_1^{(8)}$ state. }
\label{fig_feynman}
\end{figure}

In order to obtain non-vanishing $p_T$-distributions of final quarkonium we consider partonic reactions $g+g \to \qqpairr +g$. Corresponding Feynman diagrams are shown in Fig.~\ref{fig_feynman}. Differential cross-sections of these partonic processes for different quantum numbers $r$ were calculated by a number of authors \cite{Baier:1983va, Gastmans:1987be,Cho:1995vh, Cho:1995ce, Klasen:2003zn, Meijer:2007eb}. In our paper we use the cross sections provided in \cite{Meijer:2007eb}. We also use CTEQ6 partonic distribution functions (PDFs) \cite{Pumplin:2002vw} and LHAPDF interface \cite{Whalley:2005nh} to perform Monte-Carlo integration in \eqref{eq_nrqcd_main_pp}. In our calculations we set both $\alpha_S(\mu^2)$ and PDF $f_g(x;\mu^2)$ scales to $\mu^2 = M^2 + p_T^2$.

In Table~\ref{tab_asymptotic_pT} we collected asymptotic behavior of hadronic differential cross sections $d \sigma / dp_T$ of different $\qqpairr$ states in high and low $p_T$-regions. It can be seen from this table, that some of them are divergent in low $p_T$-region. This is due to $\hat t$- and $\hat u$-channel gluons in Fig.~\ref{fig_feynman} (a), (d), (e) that cause collinear singularity. It is interesting to note, that there is no such singularity in the case of $^3P_1^{(1)}$. This can be explained by Landau-Yang theorem which forbids formation of axial meson from two massless gluons: when $p_T$ tends to zero, a singularity caused by propagator canceled by vanishing effective vertex $gg^* \to{} ^3P_1^{(1)}$ since virtual gluon $g^*$ tends to mass shell. In order to avoid collinear singularities, in our work we consider quarkonium production with $p_T \gtrsim M$.

\begin{table}[h]
\begin{centering}
\begin{tabular}{|c|c|c||c|c|c|}
\hline 
 & $^{3}P_{1}^{(1,8)}$ 
 & $^{3}P_{0,2}^{(1)}$ 
 & $^{3}P_{1}^{(8)}$ 
 & $^{1}P_{1}^{(8)}$ , $^{3}P_{0,2}^{(8)}$
 & $^{3}S_{1}^{(8)}$\tabularnewline
\hline 
\hline 
$p_{T}\ll M$ 
& $\sim p_{T}$ 
& $\sim1/p_{T}$ 
& $\sim p_{T}$ 
& $\sim1/p_{T}$ 
& $\sim p_{T}$\tabularnewline
\hline 
$p_{T}\gg M$ 
& $\sim1/p_{T}^{5}$ 
& $\sim1/p_{T}^{5}$ 
& $\sim1/p_{T}^{5}$ 
& $\sim1/p_{T}^{5}$ 
& $\sim1/p_{T}^{3}$\tabularnewline
\hline 
\end{tabular}
\par\end{centering}
\caption{Asymptotic behavior of differential cross sections $d\sigma/dp_T$ of different $\qqpairr$ states in high and low $p_T$-regions.}
\label{tab_asymptotic_pT}
\end{table}

Color singlet matrix elements can be expressed via derivative of mesons's wave function at the origin:
\begin{equation}
\label{eq_npme_r20}
\left\langle 0 \left| \mathcal O^{\chi_{J}}[^3P_J^{(1)}]\right| 0\right\rangle = \frac{3}{4\pi} (2J + 1) \left|R'(0)\right|^2
\end{equation}
Numerical value of this parameter can be extracted from experimental value of $\chi$-meson's 2-photonic width:
\begin{eqnarray}
\Gamma(\chi_2\to \gamma\gamma) = \frac{1024}{45} \alpha^2  \frac{|R'(0)|^2}{M^4}
\end{eqnarray}
or from the corresponding potential model \cite{Munz:1996hb,Ebert:2003mu,Anisovich:2005jp,Wang:2009er,Li:2009nr,Hwang:2010iq}. Both methods give the following approximate value of the wave function
\begin{eqnarray}
 |R'(0)|^2 &\approx& 0.075\,\mathrm{GeV}^5.
 \label{eq:phR0}
\end{eqnarray}
Later we will refer to this value as ``phenomenological''. It should be stressed, however, that in our further studies we will consider this parameter as a free parameter and determine it from the fit.

Other non-perturbative matrix elements $\npme$ satisfy the following multiplicity relations:
\begin{equation}
\label{eq_npme_multiplicity}
\left\langle 0 \left| \mathcal O^{\chi_{J}}[\phantom{\!\!\!\!\!\!\!\!\!\!\!^3P_0^{(8)}} r]\right| 0\right\rangle = (2J + 1) \left\langle 0 \left| \mathcal O^{\chi_{0}}[\phantom{\!\!\!\!\!\!\!\!\!\!\!^3P_0^{(8)}} r]\right| 0\right\rangle, \qquad
\left\langle 0 \left| \mathcal O^{\chi_{J}}[^3P_{J'}^{(8)}]\right| 0\right\rangle =  \left\langle 0 \left| \mathcal O^{\chi_{J}}[^3P_0^{(8)}]\right| 0\right\rangle
\end{equation}
With the use of \eqref{eq_npme_r20} and \eqref{eq_npme_multiplicity} one has only 4 independent matrix elements. On the other hand, in the considered $p_T$-regions the cross sections $d\sigma/dp_T$ of $^3P_{J}^{(8)}$ and $^1P_{1}^{(8)}$ states have almost same $p_T$-dependence, so the corresponding matrix elements can be determined only in a linear combination:

\begin{multline}
\left\langle 0 \left| \mathcal O^{\chi_{0}}[^1P_1^{(8)}]\right| 0\right\rangle \frac{d \sigma\left(^1P_1^{(8)}\right)}{d p_T} +  
\sum_{J'} \left\langle 0 \left| \mathcal O^{\chi_{0}}[^3P_{J'}^{(8)}]\right| 0\right\rangle \frac{d \sigma\left(^3P_{J'}^{(8)}\right)}{d p_T} = \\
=  \frac{d \sigma\left(^1P_1^{(8)}\right)}{d p_T}  \left( 
\left\langle 0 \left| \mathcal O^{\chi_{0}}[^1P_1^{(8)}]\right| 0\right\rangle +  
\left\langle 0 \left| \mathcal O^{\chi_{0}}[^3P_{0}^{(8)}]\right| 0\right\rangle \sum_{J'}  \left(\left. \frac{d \sigma\left(^3P_{J'}^{(8)}\right)}{d p_T} \right/ \frac{d \sigma\left(^1P_1^{(8)}\right)}{d p_T}\right) 
\right) = \\ = 
\frac{d \sigma\left(^1P_1^{(8)}\right)}{d p_T}  
\left( \left\langle 0 \left| \mathcal O^{\chi_{0}}[^1P_1^{(8)}]\right| 0\right\rangle + k 
 \left\langle 0 \left| \mathcal O^{\chi_{0}}[^3P_{0}^{(8)}]\right| 0\right\rangle \right),
\end{multline}
where in the considered $p_T$ region ($4  \leq p_T \leq 30$):
\begin{equation}
k = \sum_{J'}  \left(\left. \frac{d \sigma\left(^3P_{J'}^{(8)}\right)}{d p_T} \right/ \frac{d \sigma\left(^1P_1^{(8)}\right)}{d p_T}\right) \approx 6.8 \pm 0.2.
\end{equation}
Through the rest of paper we will use the following notation for the non-perturbative parameters that we determine from fit:
\begin{equation}
\OS = \left\langle 0 \left| \mathcal O^{\chi_{0}}[^3S_1^{(8)}]\right| 0\right\rangle, 
\quad
\OP = \left\langle 0 \left| \mathcal O^{\chi_{0}}[^1P_1^{(8)}]\right| 0\right\rangle + k 
\sum_{J'} \left\langle 0 \left| \mathcal O^{\chi_{0}}[^3P_{J'}^{(8)}]\right| 0\right\rangle.
\end{equation}

In order to distinguish the impact of color octet states, let us consider $p_T$-dependence of the ratio $\chirat{J_2}{J_1}$ which can be measured more easily in experiment than the absolute cross sections. From Tab.~\ref{tab_asymptotic_pT} it is seen that $S$-wave color octet state should dominate at high $p_T$. Thus, at $p_T \gg M$ we should have 
\begin{equation}
\label{eq_chirat_cotet_assympt}
\chirat{J_2}{J_1} \approx \frac{2J_2 + 1}{2J_1 + 1},
\end{equation}
and for example in the case of $\chirat{2}{1}$ it will be 5/3. On the other hand, if we assume that contribution of $S$-wave is negligibly small at high $p_T$, than we have
\begin{eqnarray}
\chirat{2}{1} & \approx & \frac{1}{3} \,+\, \frac{\OP}{0.75\,\left|R'(0)\right|^{2}+ 0.64\, \OP },
\\
\chirat{0}{1} & \approx & \frac{1}{6} \,+\, \frac{\OP}{6\,\left|R'(0)\right|^{2} + 5.11\, \OP  },
\\
\chirat{0}{2} & \approx & \frac{1}{2} \,- \, \frac{\OP}{3.3\,\left|R'(0)\right|^{2}+ 0.56\, \OP }.
\end{eqnarray}
The first term on the r.h.s. of these expressions corresponds to pure CS model predictions.

\section{Analysis of $\chi_c$-mesons production rate in high energy hadronic collisions}
\label{sec_charm}
For a long time the only available experimental data on $p_T$-distribution of $\chi_c$ cross section were CDF measurements of $J/\psi$-mesons produced via radiative $\chi_c$ decays \cite{Abe:1997yz}. Later the CDF collaboration also measured $p_T$-dependence of $\chirat{c2}{c1}$ \cite{Abulencia:2007bra}. With the launch of the LHC a new data on $\chirat{c2}{c1}$ became available (CMS \cite{Chatrchyan:2012ub}, LHCb run1 \cite{LHCb:2012ac} and run 2 \cite{Aaij:2013dja}, ATLAS \cite{TheATLAScollaboration:2013bja}). ATLAS collaboration also measured $p_T$-distributions of absolute cross sections of $\chi_{c1}$- and $\chi_{c2}$-mesons \cite{TheATLAScollaboration:2013bja}. In order to determine non-perturbative NRQCD parameters we use $\chi^2$-criteria to fit all available data.

It should be noted, that from the experimental point of view it is easier to measure the $p_T$-dependence of ratio $\chirat{c2}{c1}$ since experimental efficiencies are canceled in this ratio and it is measured with a higher precision. As a first step of our analysis we performed fit of the available data on $\chirat{c2}{c1}$ ratio.

\begin{figure}
\centerline{\includegraphics[width=0.5\textwidth]{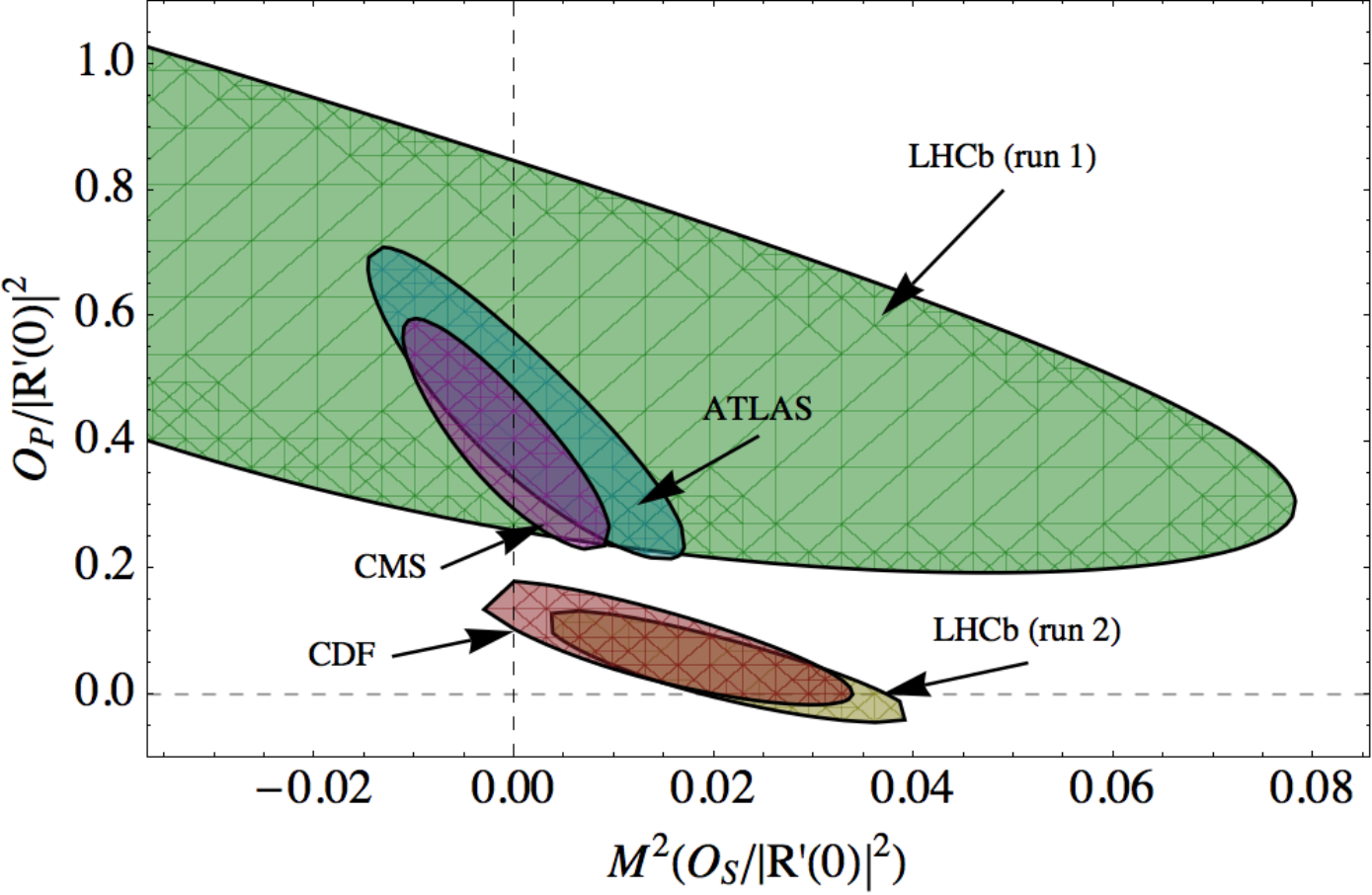}}
\caption{Acceptable regions of dimensionless parameters $M_{\chi_c}^2\OS/\RR$ and $\OP/\RR$ obtained by fitting each experiment on ratio $\chirat{2}{1}$.  }
\label{fig_regions}
\end{figure}

It is clear that only two independent parameters can be determined from the fit of the ratio $\chirat{c2}{c1}$, since one can divide both numerator and denominator by one of non-perturbative parameters. Below we will use dimensionless combinations $M_{\chi_c}^2\OS/\RR$ and $\OP/\RR$ as such free parameters. For each available experimental data set we performed $\chi^2$-fit and determined the acceptable region of non-perturbative parameters by imposing a restriction $\chi^2/DOF < \chi^2_{\min}/DOF + 1$. Fig.~\ref{fig_regions} shows the obtained acceptable regions for each experiment. It is seen from this figure, that experimental data can be separated into two groups that give two disjoint regions for non-perturbative parameters: first group --- LHCb (run 1) \cite{LHCb:2012ac}, CMS \cite{Chatrchyan:2012ub} and ATLAS \cite{TheATLAScollaboration:2013bja}, and second group ---  LHCb (run 2) \cite{Aaij:2013dja} and CDF \cite{Abulencia:2007bra}. Obviously, these two regions will merge, if we increase the errors of our estimations (by allowing a larger variation of $\chi^2$). On the other hand, such picture qualitatively shows that the octet parameters are highly sensitive to the ratio and, thus, the ratio $\chirat{c2}{c1}$ is a perfect tool to study the impact of the octet states.

\begin{figure}
\subfloat[Fit of ratio \cite{LHCb:2012ac, Chatrchyan:2012ub, TheATLAScollaboration:2013bja} and overall fit \cite{Aaij:2013dja,Abulencia:2007bra,LHCb:2012ac, Chatrchyan:2012ub, TheATLAScollaboration:2013bja}]{
	\includegraphics[width=0.5\textwidth]{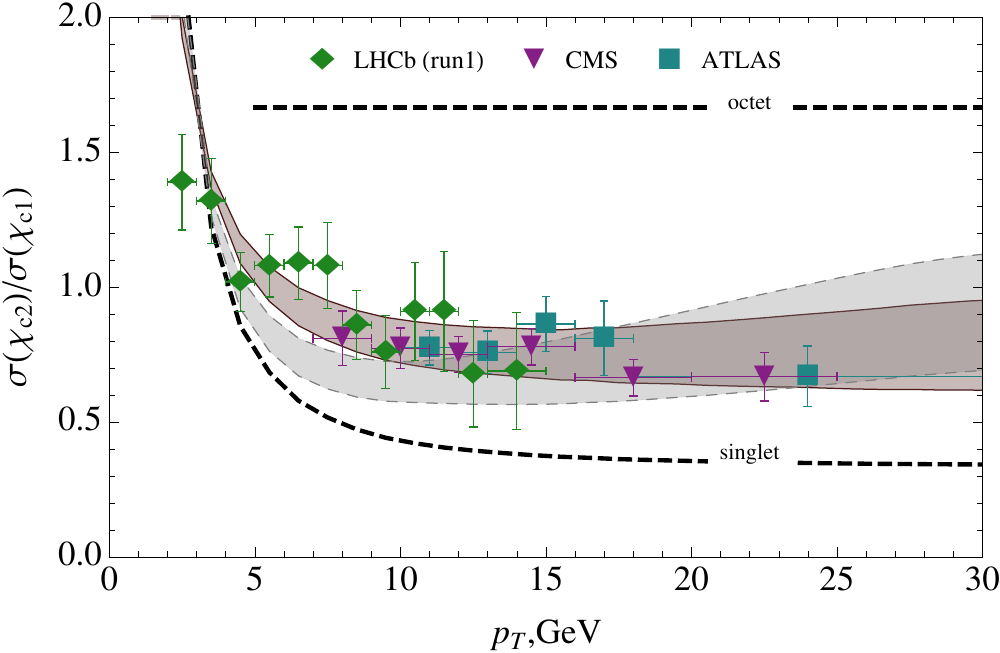} 
	\label{fig_ratio_group1}
} 
\subfloat[Fit of ratio \cite{Aaij:2013dja,Abulencia:2007bra} and overall fit \cite{Aaij:2013dja,Abulencia:2007bra,LHCb:2012ac, Chatrchyan:2012ub, TheATLAScollaboration:2013bja}]{
	\includegraphics[width=0.5\textwidth]{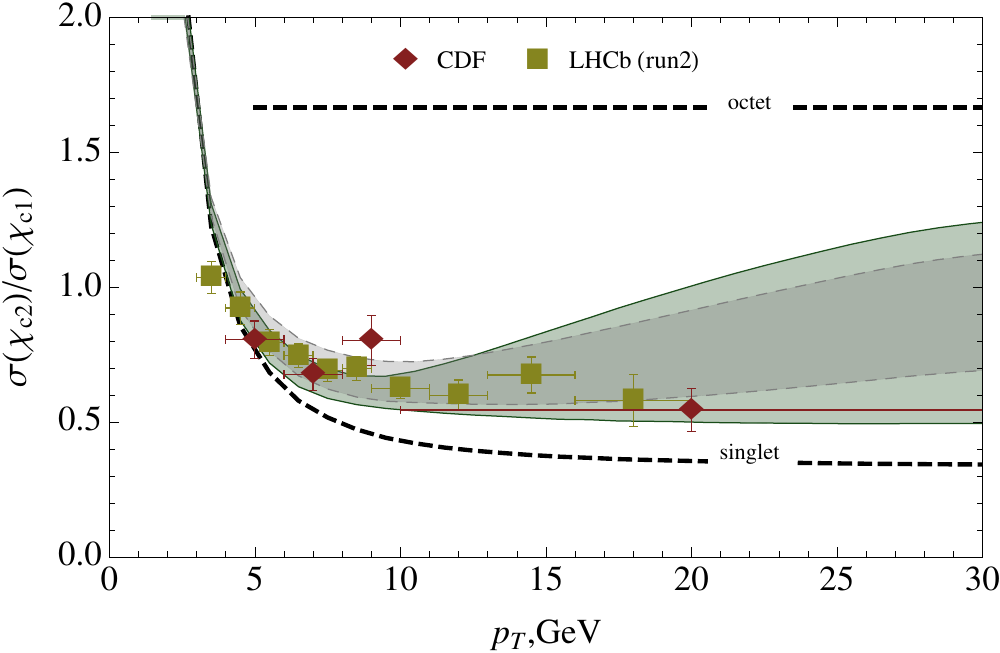}
	\label{fig_ratio_group2}
} 
\\
\subfloat[Combined fit of  ratio \cite{LHCb:2012ac, Chatrchyan:2012ub, TheATLAScollaboration:2013bja} with spectrum \cite{Abe:1997yz} and overall fit \cite{Abe:1997yz, Aaij:2013dja,Abulencia:2007bra,LHCb:2012ac, Chatrchyan:2012ub, TheATLAScollaboration:2013bja}]{
	\includegraphics[width=0.5\textwidth]{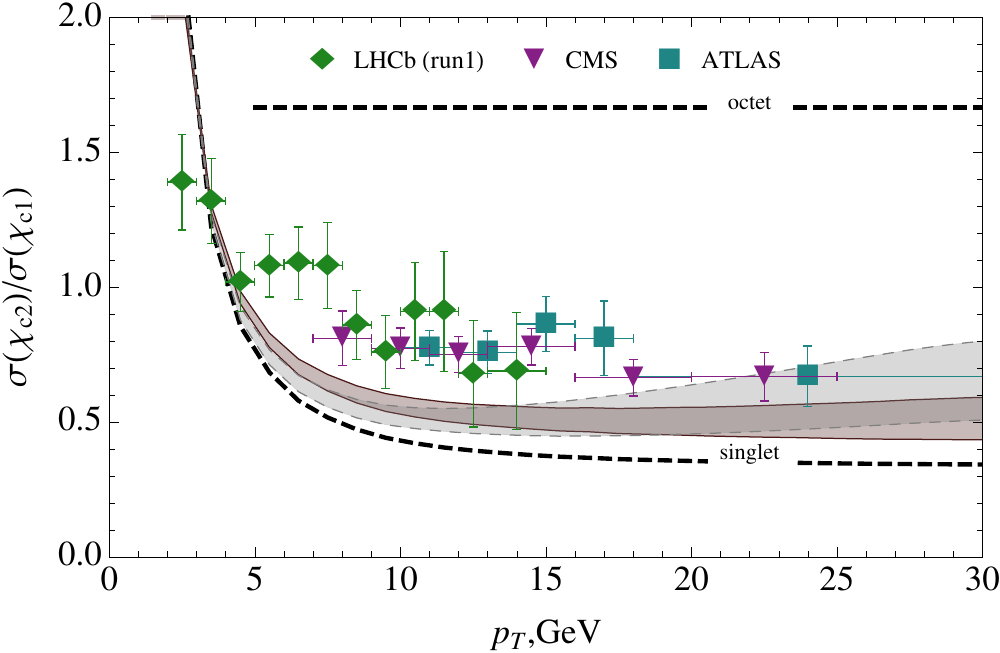}
	\label{fig_ratio_group1sp}
} 
\subfloat[Combined fit of  ratio \cite{Aaij:2013dja,Abulencia:2007bra} with spectrum \cite{Abe:1997yz} and overall fit \cite{Abe:1997yz, Aaij:2013dja,Abulencia:2007bra,LHCb:2012ac, Chatrchyan:2012ub, TheATLAScollaboration:2013bja}]{
	\includegraphics[width=0.5\textwidth]{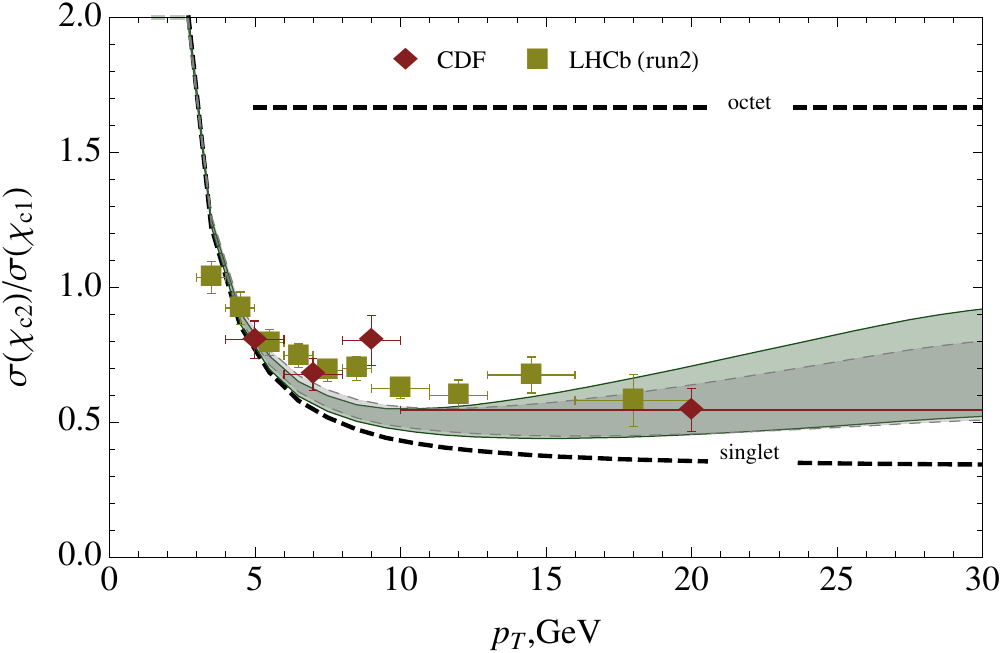}
	\label{fig_ratio_group2sp}
} 
\caption{Theoretical predictions for the ratio $\chirat{c2}{c1}$ with NRQCD parameters obtained by fitting different groups of experiments. The results of overall fit of all data shown as gray area with dashed boundaries. Upper dashed line corresponds to asymptotic NRQCD predictions for high $p_T$. Lower dashed line --- predictions of pure color singlet model.}
\label{fig_ratio_groups_sp}
\end{figure}

In order to quantitatively understand the differences in obtained parameters regions, we performed fit of each of two groups of experiments and overall fit of all data on ratio. Tab.~\ref{tab_fit_ratio} summarizes the results of these fits and Figs.~\ref{fig_ratio_group1}~and~\ref{fig_ratio_group2} illustrate the results in comparison with experimental data. One can see, that the difference in the used experimental data strongly affects on parameter values. On the other hand, from Figs.~\ref{fig_ratio_group1}~and~\ref{fig_ratio_group2} it is seen that the difference in parameter values changes the theoretical ratio not so crucially and all fits shows that CS state is dominating while the presence of $S$-wave octet raises the ratio at hight $p_T$ (according to \eqref{eq_chirat_cotet_assympt}) and the presence of $P$-wave octet constantly raises the ratio in all $p_T$ region (leaving, however, it almost parallel to pure color single predictions). 

Summarizing the above considerations, we conclude that the CS term is dominating in ratio $\chirat{c2}{c1}$ and the ratio itself is highly sensitive to the relative contributions of the octet states.

\begin{table}[h]
\centering
\begin{tabular}{|c|c|c|c|}
\hline
Experiment &$\OSRR, 10^{-2}$ & $\OPRR, 10^{-1}$ & $\chi^2/DOF$\\
\hline\hline
\cite{LHCb:2012ac,Chatrchyan:2012ub,TheATLAScollaboration:2013bja} 
&$0 \pm 0.46$ &  $4.24 \pm 0.70$ & $1.31$\\
\hline
\cite{Aaij:2013dja,Abulencia:2007bra}
 & $1.88 \pm 1.13$ & $0.48 \pm 0.06$ & $4.45$\\
\hline
all \cite{LHCb:2012ac,Chatrchyan:2012ub,TheATLAScollaboration:2013bja, Aaij:2013dja,Abulencia:2007bra}
 & $1.41 \pm 0.6$ & $1.23 \pm 0.04$ & $3.86$\\
\hline
\end{tabular}
\caption{Results of ratio fit for two groups of experiments and global fit of all data.}
\label{tab_fit_ratio}
\end{table}


The next step of our analysis is determining the whole set of non-perturbative parameters. For this purpose, we considered CDF data on $p_T$-spectrum of $J/\psi$ produced via $\chi_c$ decays in addition to data on ratio (thus we have enough degrees of freedom to determine three independent parameters). As in the previous case we separately performed fit of two sets of data on ratio ratio with CDF spectrum and an overall fit of all data (including ratio and CDF spectrum). The results of these fits are presented in Tab.~\ref{tab_fit_spectrum}. Fig.~\ref{fig_ratio_group1sp}~and~\ref{fig_ratio_group2sp} illustrates obtained theoretical ratio $\chirat{c2}{c1}$ in comparison with experimental data. It is seen, that inclusion of  spectrum data into fitting procedure reduces the acceptable range of octet parameters and worsens the agreement with experiments on ratio. At the same time, although the difference between the octet parameters still is relatively significant (Tab.~\ref{tab_fit_spectrum}), from Fig.~\ref{fig_ratio_group1sp}~and~\ref{fig_ratio_group2sp} it is seen that all fits are very close. This is because the total contribution of octet states is very small and CS term is dominating.

As it was already noted, we don't fix the value of color singlet matrix element, but determine it from the fit. It can be clearly seen from Tab.~\ref{tab_fit_spectrum}, that obtained value of CS parameter exceeds significantly the phenomenological value (\ref{eq:phR0}). The possible reason is that the very use of potential model predictions and $\chi_{c2}$ decay width for charmonium production at high energies is rather questionable. From double charmonia production in exclusive electron-positron annihilation \cite{Abe:2002rb,Braguta:2008qe,Braguta:2008hs} we know, that with the increase of the interaction energy the width of the momentum distributions of heavy quarks in quarkonia also increases. In coordinate space it corresponds to the increase of the charmonium wave function and its derivative at the origin.

It is also necessary to point out that our result for $S$-wave color octet matrix element is approximately an order of magnitude less than the results of another fits (see e.g. \cite{Kniehl:2003pc}). On the one hand this is because a recent data on $\chirat{2}{1}$ strongly shows that this term is suppressed. On the other hand this is also because the value of CS parameter in our work is determined from fit and is significantly greater then phenomenological value used in other works; thus CS term is almost enough for explaining $p_T$-spectrum and octet state is required only for description of $\chirat{2}{1}$ ratio. 

\begin{figure}
\subfloat[Total cross section.]{	\includegraphics[width=0.5\textwidth]{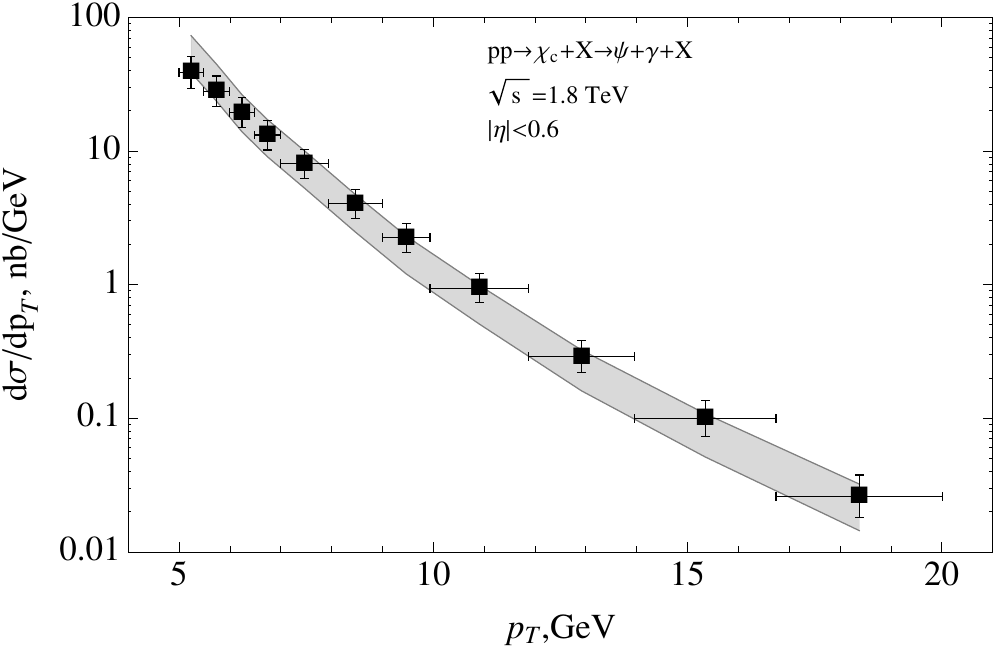}\label{fig_cdf_spectrum_total}} 
\subfloat[Contribution of different NRQCD states.]{	\includegraphics[width=0.5\textwidth]{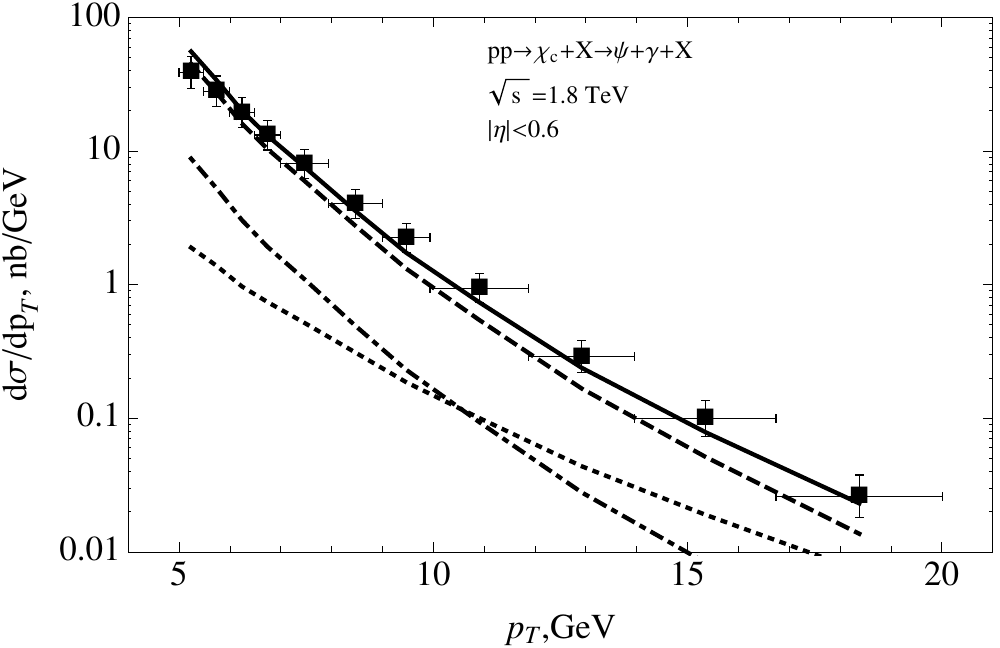}\label{fig_cdf_spectrum_term}} 
\caption{Theoretical predictions for the cross section of $J/\psi$-mesons produced via radiative $\chi_c$ decays with NRQCD parameters obtained by fitting a whole set of data on ratio \cite{LHCb:2012ac, Chatrchyan:2012ub, TheATLAScollaboration:2013bja, Aaij:2013dja, Abulencia:2007bra} and CDF data on spectrum \cite{Abe:1997yz}. CDF data shown with filled squares. The left figure shows the total cross section with uncertainties, right --- contribution of different NRQCD state using best fit parameters. Solid line --- sum over all states, dashed ---- color singlet, dot-dashed --- $P$-wave octet, dotted --- $S$-wave octet. 
}
\label{fig_cdf_spectrum}
\end{figure}

Fig.~\ref{fig_cdf_spectrum} shows the results of global fit in comparison with CDF points. From Fig.~\ref{fig_cdf_spectrum_term} it is clear that dominant contribution is due to color singlet quarkonium production. It is important to note that one will obtain nearly the same total curve if use the parameters from first or second column of Tab.~\ref{tab_fit_spectrum}; this is a consequence of the small role of octet contributions to the total cross section


\begin{table}[h]
\centering
\begin{tabular}{|c|c|c|c|}
\hline
 & \cite{LHCb:2012ac,Chatrchyan:2012ub,TheATLAScollaboration:2013bja} and \cite{Abe:1997yz}
 & \cite{Aaij:2013dja,Abulencia:2007bra} and \cite{Abe:1997yz} 
 & all \cite{LHCb:2012ac, Chatrchyan:2012ub, TheATLAScollaboration:2013bja, Aaij:2013dja, Abulencia:2007bra, Abe:1997yz} \\
 \hline \hline
 $\chi^2/DOF$ & 1.24 & 2.64 & 3.14 \\
 \hline
 $\RR, \mbox{GeV}^{5}$ & $0.27\pm 0.03$ & $0.38\pm0.05$ & $0.35\pm0.05$ \\
 \hline
 $\OS, 10^{-3}\,\mbox{GeV}^{3}$ & $0.\pm0.1$ & $0.66\pm0.27$ & $0.44\pm0.16$\\
 \hline
 $\OP, 10^{-1}\,\mbox{GeV}^{5}$ & $1.14\pm0.16$ & $0.14\pm0.15$ & $0.41\pm0.14$\\
 \hline
\end{tabular}
\caption{Results of fitting both ratio and spectrum for two groups of experiments and for all data.}
\label{tab_fit_spectrum}
\end{table}


\begin{figure}
\subfloat[Production of $\chi_{c1}$.]{	\includegraphics[width=0.5\textwidth]{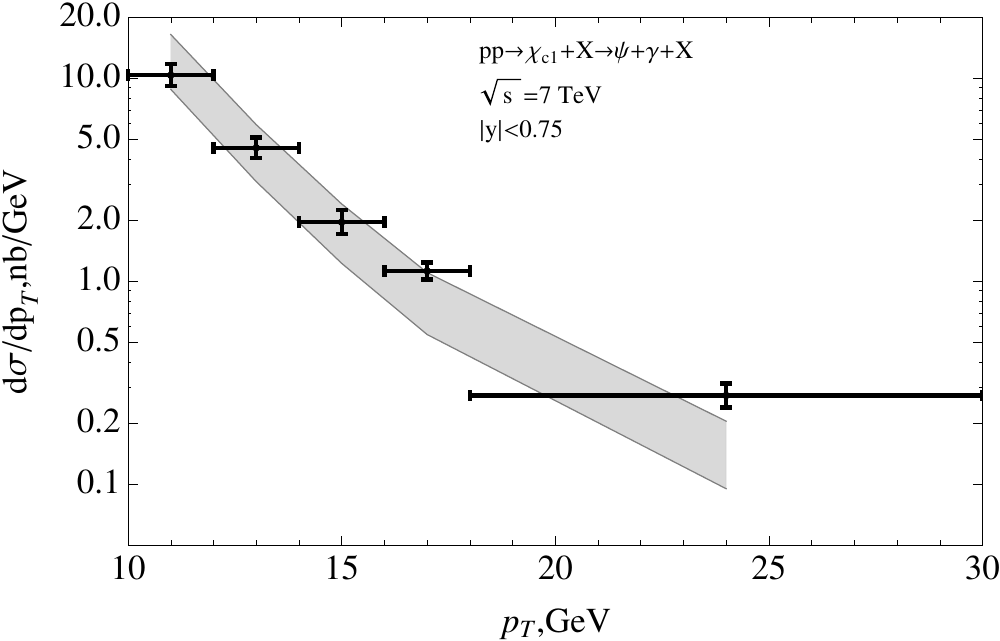}\label{fig_atlas1_spectrum}} 
\subfloat[Production of $\chi_{c2}$.]{	\includegraphics[width=0.5\textwidth]{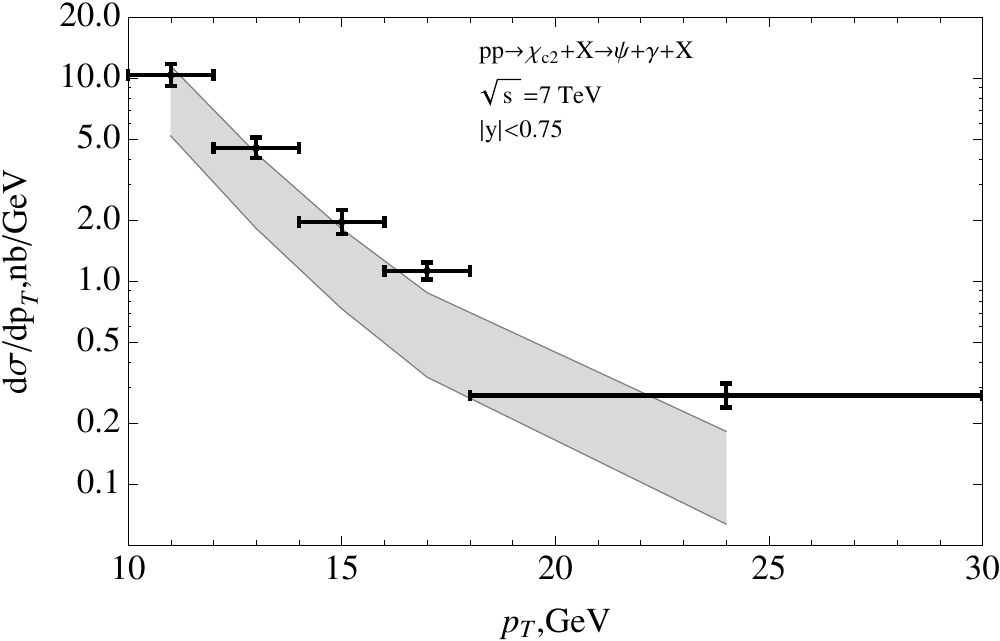}\label{fig_atlas2_spectrum}} 
\caption{Theoretical predictions for the cross sections of prompt $\chi_{c1,2}$-mesons with NRQCD parameters obtained by fitting a whole set of data on ratio \cite{LHCb:2012ac, Chatrchyan:2012ub, TheATLAScollaboration:2013bja, Aaij:2013dja, Abulencia:2007bra} and CDF data on spectrum \cite{Abe:1997yz}. ATLAS data on spectrum \cite{TheATLAScollaboration:2013bja} (not included in fit) shown with points. }
\label{fig_atlas_spectrum}
\end{figure}

Figs.~\ref{fig_atlas1_spectrum} and \ref{fig_atlas2_spectrum} show theoretical predictions for $\chi_{c1}$ and $\chi_{c2}$ cross sections in comparison with ATLAS points \cite{TheATLAScollaboration:2013bja} (which are not included in fit). As one can see, there is a good agreement between theory and experiment.

\section{Prediction for $\chi_b$-mesons production rates}
\label{sec_bottom}
In this section we discuss $\chi_b$-production in high energy hadronic collisions. Unfortunately, there is very little data on $\chi_b$-production cross sections, only CMS measurement of $p_T$-dependence of $\chirat{b2}{b1}$ \cite{CMS:2013yha} is available. The lack of experimental data does not allow us to determine NRQCD parameters from fit of the data. So, in order to make predictions on $\chi_b$ cross sections we will use NRQCD scaling rules.

From the dimension analysis and NRQCD velocity scaling rules the following relations can be obtained:
\begin{eqnarray}
\label{eq_nrqcd_scaling}
M_{\chi_c}^2 \frac{\OSp{c}}{\RRp{c}} \approx M_{\chi_b}^2 \frac{\OSp{b}}{\RRp{b}},
\quad
\frac{1}{v_{\chi_c}^2}\frac{\OPp{c}}{\RRp{c}} \approx \frac{1}{v_{\chi_b}^2} \frac{\OPp{b}}{\RRp{b}},
\end{eqnarray}
where we omitted the dependence of $\chi_b$ parameters on radial quantum number $n$, since this dependence cancels in the ratio. It can be shown \cite{Likhoded:2012hw} that if we neglect the dependence on $v$, then the ratio for $\chi_b$-mesons can be obtained from the ratio for $\chi_c$-mesons by using a simple scaling relation:
\begin{equation}
\label{eq_scaleRatioApprox}
\left. \frac{d\sigma_{b2}(zp_T;s)}{dp_T}\right/\frac{d\sigma_{b1}(zp_T;s)}{dp_T}
= 
\left. \frac{d\sigma_{c2}(p_T;s)}{dp_T}\right/\frac{d\sigma_{c1}(p_T;s)}{dp_T}, \quad \mbox{where} \quad z = M_{\chi_b}\left/M_{\chi_c}\right. .
\end{equation}
This simple fact relies only on the assumption that ratio $\chirat{2}{1}$ depends only on three dimensional parameters:  \(d \sigma_J/d p_T \equiv d \sigma_J/d p_T(s,p_T,M)\). As it seen from \eqref{eq_nrqcd_scaling}, in the case of NRQCD, equation \eqref{eq_scaleRatioApprox} is violated by $O(v^2)$ octet $P$-wave terms from quarkonium Fock space \eqref{eq_fock_structure}. On the other hand, scaling \eqref{eq_scaleRatioApprox} is still valid if the latter states are highly suppressed.

\begin{figure}
\centering
\subfloat[Parameters from fitting \cite{Abe:1997yz, LHCb:2012ac, Chatrchyan:2012ub, TheATLAScollaboration:2013bja} and NRQCD scaling]{\includegraphics[width=0.5\textwidth]{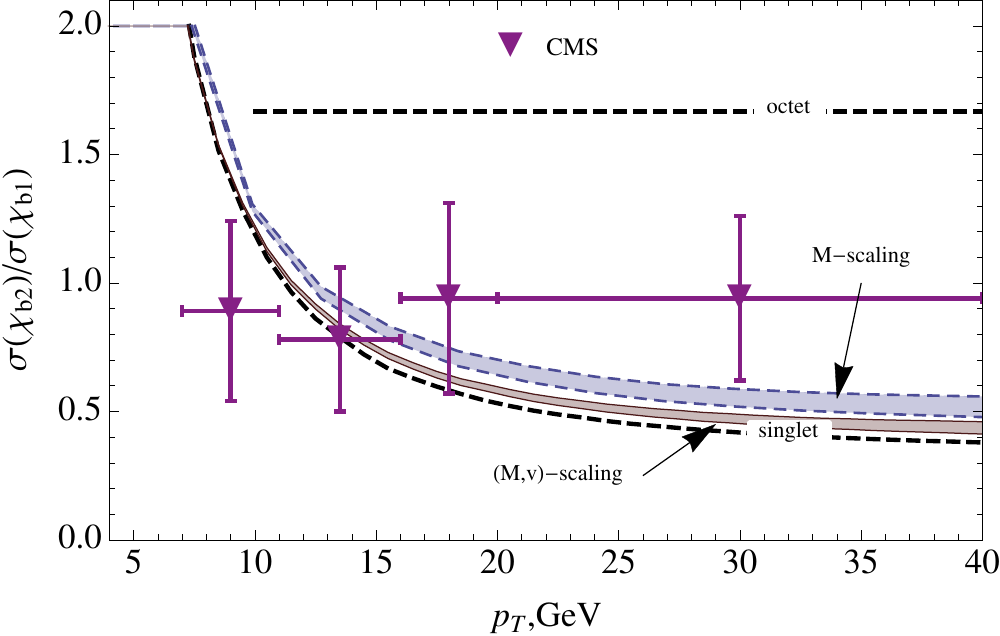}} 
\subfloat[Parameters from fitting \cite{Abe:1997yz, Aaij:2013dja,Abulencia:2007bra} and NRQCD scaling]{\includegraphics[width=0.5\textwidth]{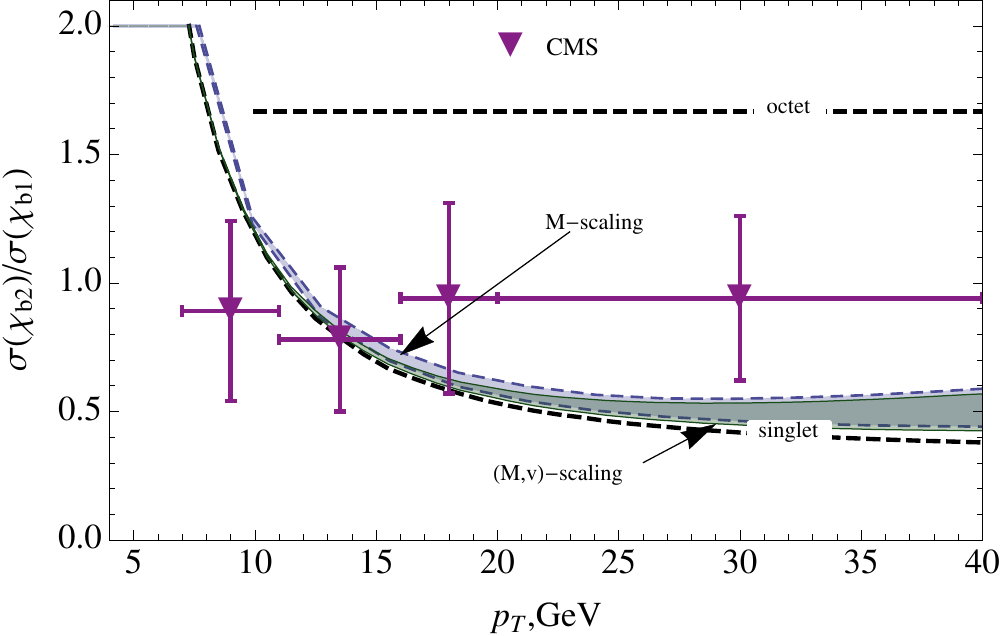}} 
\caption{Predictions of $p_T$-dependence of $\chi_b$-mesons ratio $\chirat{b2}{b1}$. Non-perturbative NRQCD parameters for $\chi_b$-mesons obtained from Tab.~\ref{tab_fit_spectrum} using NRQCD velocity scaling rules \eqref{eq_nrqcd_scaling} (labeled as (M,v)-scaling) and using simple scaling \eqref{eq_scaleRatioApprox} (labeled as M-scaling). CMS points are taken from \cite{CMS:2013yha}. Upper dashed line corresponds to asymptotic NRQCD predictions for high $p_T$. Lower dashed line --- predictions of pure color singlet model.}
\label{fig_ratio_chi_b}
\end{figure}

Fig.~\ref{fig_ratio_chi_b} shows our predictions for $\chirat{b2}{b1}$ ratio with NRQCD  parameters for $\chi_b$-mesons obtained from Tab.~\ref{tab_fit_spectrum} using NRQCD velocity scaling rules \eqref{eq_nrqcd_scaling} (labeled as (M,v)-scaling) and using simple scaling \eqref{eq_scaleRatioApprox} (labeled as M-scaling). It is clear from this figure, that predictions obtained via \eqref{eq_nrqcd_scaling} and \eqref{eq_scaleRatioApprox} give approximately same result. Moreover, both groups of NRQCD parameters (obtained from fitting two groups of data) also give approximately same predictions and $\chi_b$-ratio is very close to color singlet predictions. This is because according to \eqref{eq_nrqcd_scaling} all octet states are more suppressed for bottomonium. Predictions for bottomonium obtained using parameters from overall fit (last column in Tab.~\ref{tab_fit_spectrum}) are negligibly different from right picture on Fig.~\ref{fig_ratio_chi_b}.

It is interesting to note, that both $\chi_c$ and $\chi_b$ ratios $\chirat{2}{1}$ increase in low $p_T$-region. In the case of charmonium this fact is not so notable since, as it was discussed in Sec.~\ref{sec_theory}, $\chi_2$ has collinear singularity at low $p_T$. On the other hand, in the case of bottomonium such behavior observed for relatively large $p_T$ ($10 \, \mbox{GeV} \lesssim p_T \lesssim 15\, \mbox{GeV}$) and cannot be considered as a consequence of collinear divergence only. In the case of color singlet $^3P_1$ state Landau-Yang theorem leads to a cancellation of $\hat t$- and $\hat u$-channel gluon propagators by a vertex $g^* g^* \to ^3P_1^{(1)}$ which vanishes when gluons tend to mass shell (with $p_T \to 0$). Since color singlet term dominates and there is no such cancellation for tensor state, this explains the increasing of ratio at low $p_T$. It can be assumed that in the higher $\alpha_S$ orders such behavior will be still valid since such cancellation will always reduce power of $p_T$ in denominator of $^3P_1^{(1)}$ state cross section.

\begin{figure}
\centering
\subfloat[$\chi_{b1}$ for LHCb]{\includegraphics[width=0.5\textwidth]{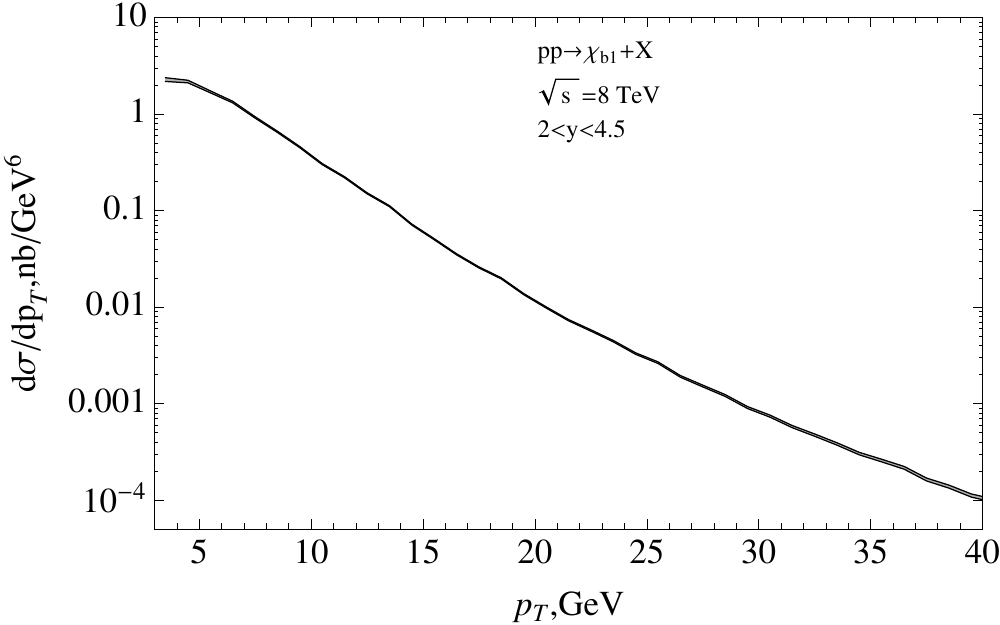}} 
\subfloat[$\chi_{b2}$ for LHCb]{\includegraphics[width=0.5\textwidth]{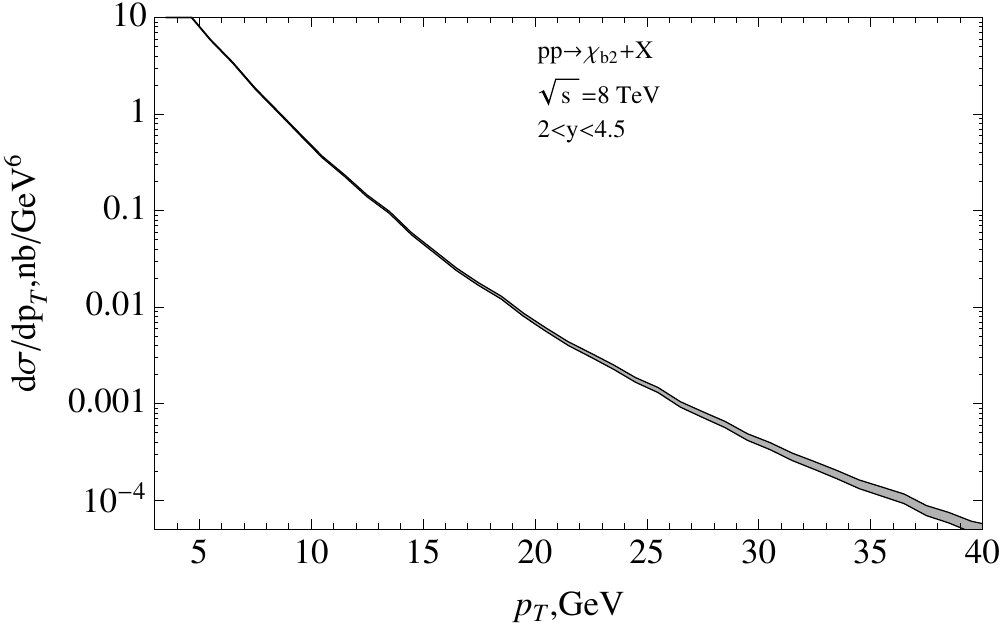}} 
\\
\subfloat[$\chi_{b1}$ for ATLAS]{\includegraphics[width=0.5\textwidth]{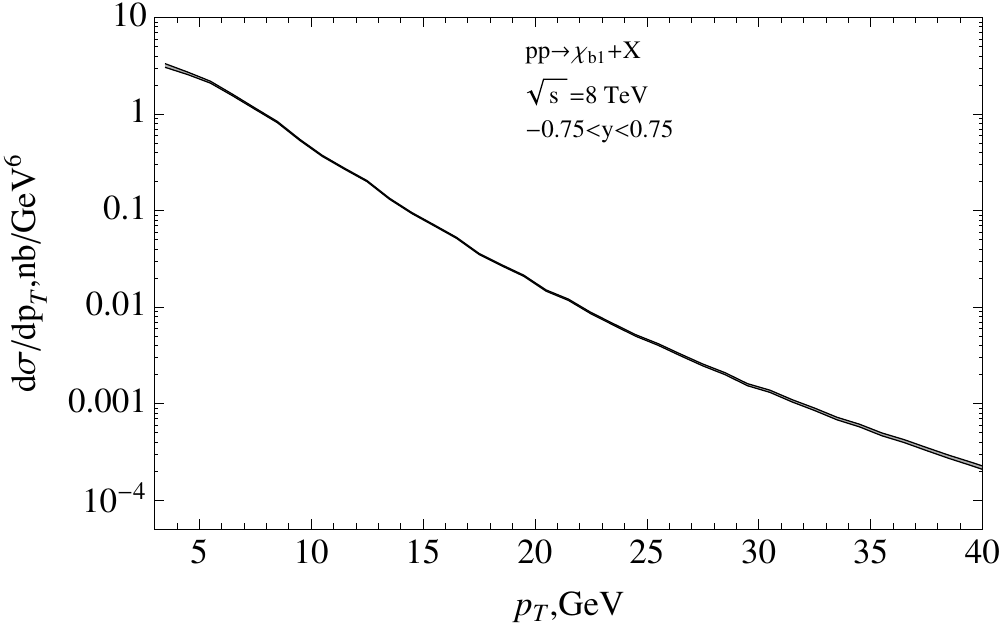}} 
\subfloat[$\chi_{b2}$ for ATLAS]{\includegraphics[width=0.5\textwidth]{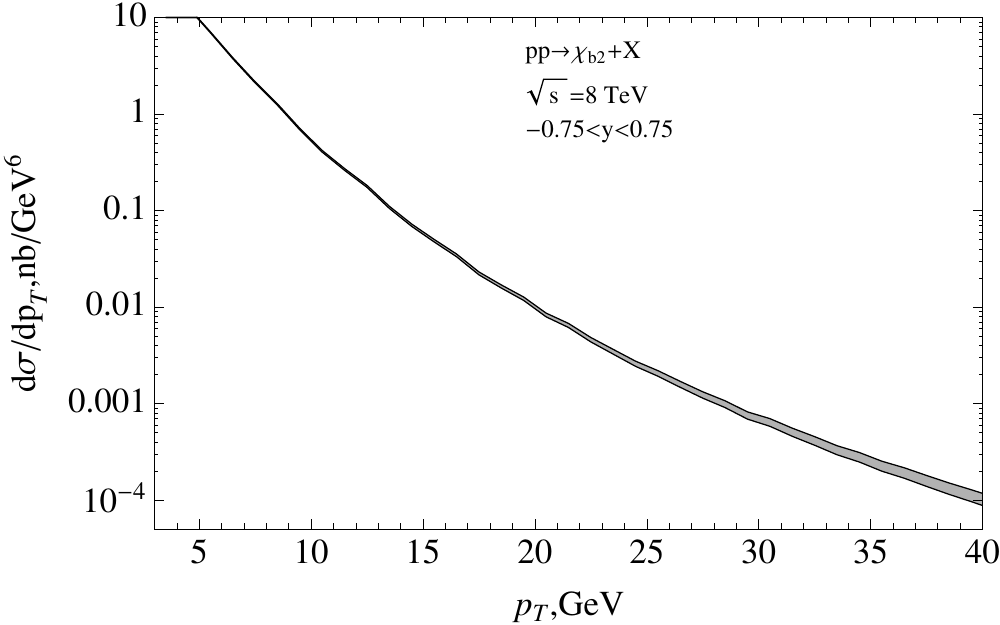}} 
\caption{Predictions of $\chi_{b1,2}$ cross sections (divided by $\RRp{b}$) for different kinematic regions. Non-perturbative NRQCD parameters obtained using NRQCD velocity scaling rules \eqref{eq_nrqcd_scaling} and parameters from global fit presented in Tab.~\ref{tab_fit_spectrum}.}
\label{fig_spectrum_chi_b}
\end{figure}

Fig.~\ref{fig_spectrum_chi_b} shows predictions of $\chi_{b1}$- and $\chi_{b2}$ production cross sections in different kinematical regions. Presented cross sections are divided by $\chi_b$-wave function $\RRp{b}$, since the latter can not be determined using only NRQCD scaling rules. One can find a table with $\RRp{b}$ values obtained in different potential models in Ref.~\cite{Likhoded:2012hw}.

\section{Conclusions}
\label{sec_conclusions}
In the present paper we have considered production of $\chi_{c}$- and $\chi_{b}$-mesons in high energy hadronic collisions in the framework of NRQCD. In order to obtain $p_T$-distributions we considered partonic processes $g + g \to \chi_{c,b} + g$. The non-relativistic nature of quarkonium bound state provides a small parameter in a model --- relative velocity $v$ of $\qqpair$ pair. NRQCD allows one to organize different terms in quarkonium Fock space in terms of $v$. In our consideration we used both leading terms on $v$ and terms of the relative order $O(v^2)$. The latter terms should give a noticeable contribution to the cross sections since $v^2$ is about 0.3 for charmonium and 0.1 for bottomonium.

The main part of our paper is devoted to study of $\chi_c$ meson production. According to our analysis the fit of all available data shows, that color singlet components give dominant contributions to $\chi_{cJ}$ mesons production cross sections, while the contributions of color octet components can safely be neglected in description of $p_T$ distributions of individual charmonium mesons. The ratios of the differential cross sections (e.g. $d\sigma(\chi_{c2})/d\sigma(\chi_{c1})$, on the other hand, are extremely sensitive to values of color octet parameters. It is shown in our paper that combined fit of available experimental data allows one to determine all model parameters and achieve a good agreement between theoretical predictions and experimental results. It turns out also, that obtained from the fit value of color singlet parameter $R'(0)|^2$ exceeds significantly its phenomenological estimate.

In the last section we considered production of bottomonium mesons. In order to determine the NRQCD matrix elements for bottomonium, we used NRQCD velocity scaling rules. Since the $\chi_b$ is about 3 times heavier and its relative quarks velocity $v^2$ is about 0.1, the impact of octet states on $\chi_b$-production is much weaker than in the case of charmonium. The predictions for the $p_T$-dependence of ratio $\chirat{b2}{b1}$ and absolute cross sections of $\chi_{b1,2}$ in different kinematical regions were also presented.

The author would like to thank I. Belyaev and E. Tournefier for fruitful discussions. The work was financially supported by RFBR (\#14-02-00096 A) and  grant of SAEC ``Rosatom'' and Helmholtz Association.

\section*{Bibliography}
\bibliographystyle{apsrev4-1}
\bibliography{references}

\begin{thebibliography}{29}%
\makeatletter
\providecommand \@ifxundefined [1]{%
 \@ifx{#1\undefined}
}%
\providecommand \@ifnum [1]{%
 \ifnum #1\expandafter \@firstoftwo
 \else \expandafter \@secondoftwo
 \fi
}%
\providecommand \@ifx [1]{%
 \ifx #1\expandafter \@firstoftwo
 \else \expandafter \@secondoftwo
 \fi
}%
\providecommand \natexlab [1]{#1}%
\providecommand \enquote  [1]{``#1''}%
\providecommand \bibnamefont  [1]{#1}%
\providecommand \bibfnamefont [1]{#1}%
\providecommand \citenamefont [1]{#1}%
\providecommand \href@noop [0]{\@secondoftwo}%
\providecommand \href [0]{\begingroup \@sanitize@url \@href}%
\providecommand \@href[1]{\@@startlink{#1}\@@href}%
\providecommand \@@href[1]{\endgroup#1\@@endlink}%
\providecommand \@sanitize@url [0]{\catcode `\\12\catcode `\$12\catcode
  `\&12\catcode `\#12\catcode `\^12\catcode `\_12\catcode `\%12\relax}%
\providecommand \@@startlink[1]{}%
\providecommand \@@endlink[0]{}%
\providecommand \url  [0]{\begingroup\@sanitize@url \@url }%
\providecommand \@url [1]{\endgroup\@href {#1}{\urlprefix }}%
\providecommand \urlprefix  [0]{URL }%
\providecommand \Eprint [0]{\href }%
\providecommand \doibase [0]{http://dx.doi.org/}%
\providecommand \selectlanguage [0]{\@gobble}%
\providecommand \bibinfo  [0]{\@secondoftwo}%
\providecommand \bibfield  [0]{\@secondoftwo}%
\providecommand \translation [1]{[#1]}%
\providecommand \BibitemOpen [0]{}%
\providecommand \bibitemStop [0]{}%
\providecommand \bibitemNoStop [0]{.\EOS\space}%
\providecommand \EOS [0]{\spacefactor3000\relax}%
\providecommand \BibitemShut  [1]{\csname bibitem#1\endcsname}%
\let\auto@bib@innerbib\@empty
\bibitem [{\citenamefont {Kartvelishvili}\ \emph {et~al.}(1981)\citenamefont
  {Kartvelishvili}, \citenamefont {Likhoded},\ and\ \citenamefont
  {Slabospitsky}}]{Kartvelishvili:1980uz}%
  \BibitemOpen
  \bibfield  {author} {\bibinfo {author} {\bibfnamefont {V.}~\bibnamefont
  {Kartvelishvili}}, \bibinfo {author} {\bibfnamefont {A.}~\bibnamefont
  {Likhoded}}, \ and\ \bibinfo {author} {\bibfnamefont {S.}~\bibnamefont
  {Slabospitsky}},\ }\href@noop {} {\bibfield  {journal} {\bibinfo  {journal}
  {Sov.J.Nucl.Phys.}\ }\textbf {\bibinfo {volume} {33}},\ \bibinfo {pages}
  {434} (\bibinfo {year} {1981})}\BibitemShut {NoStop}%
\bibitem [{\citenamefont {{Baier, R. and Ruckl, R.}}(1983)}]{Baier:1983va}%
  \BibitemOpen
  \bibfield  {author} {\bibinfo {author} {\bibnamefont {{Baier, R. and Ruckl,
  R.}}},\ }\href {\doibase 10.1007/BF01572254} {\bibfield  {journal} {\bibinfo
  {journal} {Z.Phys.}\ }\textbf {\bibinfo {volume} {C19}},\ \bibinfo {pages}
  {251} (\bibinfo {year} {1983})}\BibitemShut {NoStop}%
\bibitem [{\citenamefont {Bodwin}\ \emph {et~al.}(1995)\citenamefont {Bodwin},
  \citenamefont {Braaten},\ and\ \citenamefont {Lepage}}]{Bodwin:1994jh}%
  \BibitemOpen
  \bibfield  {author} {\bibinfo {author} {\bibfnamefont {G.~T.}\ \bibnamefont
  {Bodwin}}, \bibinfo {author} {\bibfnamefont {E.}~\bibnamefont {Braaten}}, \
  and\ \bibinfo {author} {\bibfnamefont {G.~P.}\ \bibnamefont {Lepage}},\
  }\href {\doibase 10.1103/PhysRevD.55.5853, 10.1103/PhysRevD.51.1125}
  {\bibfield  {journal} {\bibinfo  {journal} {Phys.Rev.}\ }\textbf {\bibinfo
  {volume} {D51}},\ \bibinfo {pages} {1125} (\bibinfo {year} {1995})},\ \Eprint
  {http://arxiv.org/abs/hep-ph/9407339} {arXiv:hep-ph/9407339 [hep-ph]}
  \BibitemShut {NoStop}%
\bibitem [{\citenamefont {Likhoded}\ \emph {et~al.}(2012)\citenamefont
  {Likhoded}, \citenamefont {Luchinsky},\ and\ \citenamefont
  {Poslavsky}}]{Likhoded:2012hw}%
  \BibitemOpen
  \bibfield  {author} {\bibinfo {author} {\bibfnamefont {A.K.}~\bibnamefont
  {Likhoded}}, \bibinfo {author} {\bibfnamefont {A.V.}~\bibnamefont {Luchinsky}},
  \ and\ \bibinfo {author} {\bibfnamefont {S.V.}~\bibnamefont {Poslavsky}},\
  }\href {\doibase 10.1103/PhysRevD.86.074027} {\bibfield  {journal} {\bibinfo
  {journal} {Phys.Rev.}\ }\textbf {\bibinfo {volume} {D86}},\ \bibinfo {pages}
  {074027} (\bibinfo {year} {2012})},\ \Eprint {http://arxiv.org/abs/1203.4893}
  {arXiv:1203.4893 [hep-ph]} \BibitemShut {NoStop}%
\bibitem [{\citenamefont {Likhoded}\ \emph {et~al.}(2013)\citenamefont
  {Likhoded}, \citenamefont {Luchinsky},\ and\ \citenamefont
  {Poslavsky}}]{Likhoded:2013aya}%
  \BibitemOpen
  \bibfield  {author} {\bibinfo {author} {\bibfnamefont {A.}~\bibnamefont
  {Likhoded}}, \bibinfo {author} {\bibfnamefont {A.}~\bibnamefont {Luchinsky}},
  \ and\ \bibinfo {author} {\bibfnamefont {S.}~\bibnamefont {Poslavsky}},\
  }\href@noop {} {\  (\bibinfo {year} {2013})},\ \Eprint
  {http://arxiv.org/abs/1305.2389} {arXiv:1305.2389 [hep-ph]} \BibitemShut
  {NoStop}%
\bibitem [{\citenamefont {Gastmans}\ \emph {et~al.}(1987)\citenamefont
  {Gastmans}, \citenamefont {Troost},\ and\ \citenamefont
  {Wu}}]{Gastmans:1987be}%
  \BibitemOpen
  \bibfield  {author} {\bibinfo {author} {\bibfnamefont {R.}~\bibnamefont
  {Gastmans}}, \bibinfo {author} {\bibfnamefont {W.}~\bibnamefont {Troost}}, \
  and\ \bibinfo {author} {\bibfnamefont {T.~T.}\ \bibnamefont {Wu}},\ }\href
  {\doibase 10.1016/0550-3213(87)90493-7} {\bibfield  {journal} {\bibinfo
  {journal} {Nucl.Phys.}\ }\textbf {\bibinfo {volume} {B291}},\ \bibinfo
  {pages} {731} (\bibinfo {year} {1987})}\BibitemShut {NoStop}%
\bibitem [{\citenamefont {{Cho, Peter L. and Leibovich, Adam
  K.}}(1996{\natexlab{a}})}]{Cho:1995vh}%
  \BibitemOpen
  \bibfield  {author} {\bibinfo {author} {\bibnamefont {P.}~\bibnamefont {Cho}}, \bibinfo {author} {\bibnamefont {A.K.}~\bibnamefont {Leibovich}} }\href {\doibase 10.1103/PhysRevD.53.150} {\bibfield
  {journal} {\bibinfo  {journal} {Phys.Rev.}\ }\textbf {\bibinfo {volume}
  {D53}},\ \bibinfo {pages} {150} (\bibinfo {year} {1996}{\natexlab{a}})},\
  \Eprint {http://arxiv.org/abs/hep-ph/9505329} {arXiv:hep-ph/9505329 [hep-ph]}
  \BibitemShut {NoStop}%
\bibitem [{\citenamefont {{Cho, Peter L. and Leibovich, Adam
  K.}}(1996{\natexlab{b}})}]{Cho:1995ce}%
  \BibitemOpen
  \bibfield  {author} {\bibinfo {author} {\bibnamefont {P.}~\bibnamefont {Cho}}, \bibinfo {author} {\bibnamefont {A.K.}~\bibnamefont {Leibovich}} }\href {\doibase 10.1103/PhysRevD.53.6203} {\bibfield
   {journal} {\bibinfo  {journal} {Phys.Rev.}\ }\textbf {\bibinfo {volume}
  {D53}},\ \bibinfo {pages} {6203} (\bibinfo {year} {1996}{\natexlab{b}})},\
  \Eprint {http://arxiv.org/abs/hep-ph/9511315} {arXiv:hep-ph/9511315 [hep-ph]}
  \BibitemShut {NoStop}%
\bibitem [{\citenamefont {Klasen}\ \emph {et~al.}(2003)\citenamefont {Klasen},
  \citenamefont {Kniehl}, \citenamefont {Mihaila},\ and\ \citenamefont
  {Steinhauser}}]{Klasen:2003zn}%
  \BibitemOpen
  \bibfield  {author} {\bibinfo {author} {\bibfnamefont {M.}~\bibnamefont
  {Klasen}}, \bibinfo {author} {\bibfnamefont {B.A.}~\bibnamefont {Kniehl}},
  \bibinfo {author} {\bibfnamefont {L.N.}~\bibnamefont {Mihaila}}, \ and\
  \bibinfo {author} {\bibfnamefont {M.}~\bibnamefont {Steinhauser}},\ }\href
  {\doibase 10.1103/PhysRevD.68.034017} {\bibfield  {journal} {\bibinfo
  {journal} {Phys.Rev.}\ }\textbf {\bibinfo {volume} {D68}},\ \bibinfo {pages}
  {034017} (\bibinfo {year} {2003})},\ \Eprint
  {http://arxiv.org/abs/hep-ph/0306080} {arXiv:hep-ph/0306080 [hep-ph]}
  \BibitemShut {NoStop}%
\bibitem [{\citenamefont {Meijer}\ \emph {et~al.}(2008)\citenamefont {Meijer},
  \citenamefont {Smith},\ and\ \citenamefont {van Neerven}}]{Meijer:2007eb}%
  \BibitemOpen
  \bibfield  {author} {\bibinfo {author} {\bibfnamefont {M.M.}~\bibnamefont
  {Meijer}}, \bibinfo {author} {\bibfnamefont {J.}~\bibnamefont {Smith}}, \
  and\ \bibinfo {author} {\bibfnamefont {W.L.}~\bibnamefont {van Neerven}},\
  }\href {\doibase 10.1103/PhysRevD.77.034014} {\bibfield  {journal} {\bibinfo
  {journal} {Phys.Rev.}\ }\textbf {\bibinfo {volume} {D77}},\ \bibinfo {pages}
  {034014} (\bibinfo {year} {2008})},\ \Eprint {http://arxiv.org/abs/0710.3090}
  {arXiv:0710.3090 [hep-ph]} \BibitemShut {NoStop}%
\bibitem [{\citenamefont {Pumplin}\ \emph {et~al.}(2002)\citenamefont
  {Pumplin}, \citenamefont {Stump}, \citenamefont {Huston}, \citenamefont
  {Lai}, \citenamefont {Nadolsky} \emph {et~al.}}]{Pumplin:2002vw}%
  \BibitemOpen
  \bibfield  {author} {\bibinfo {author} {\bibfnamefont {J.}~\bibnamefont
  {Pumplin}}, \bibinfo {author} {\bibfnamefont {D.}~\bibnamefont {Stump}},
  \bibinfo {author} {\bibfnamefont {J.}~\bibnamefont {Huston}}, \bibinfo
  {author} {\bibfnamefont {H.}~\bibnamefont {Lai}}, \bibinfo {author}
  {\bibfnamefont {P.~M.}\ \bibnamefont {Nadolsky}},  \emph {et~al.},\ }\href
  {\doibase 10.1088/1126-6708/2002/07/012} {\bibfield  {journal} {\bibinfo
  {journal} {JHEP}\ }\textbf {\bibinfo {volume} {0207}},\ \bibinfo {pages}
  {012} (\bibinfo {year} {2002})},\ \Eprint
  {http://arxiv.org/abs/hep-ph/0201195} {arXiv:hep-ph/0201195 [hep-ph]}
  \BibitemShut {NoStop}%
\bibitem [{\citenamefont {Whalley}\ \emph {et~al.}(2005)\citenamefont
  {Whalley}, \citenamefont {Bourilkov},\ and\ \citenamefont
  {Group}}]{Whalley:2005nh}%
  \BibitemOpen
  \bibfield  {author} {\bibinfo {author} {\bibfnamefont {M.}~\bibnamefont
  {Whalley}}, \bibinfo {author} {\bibfnamefont {D.}~\bibnamefont {Bourilkov}},
  \ and\ \bibinfo {author} {\bibfnamefont {R.}~\bibnamefont {Group}},\
  }\href@noop {} {\  (\bibinfo {year} {2005})},\ \Eprint
  {http://arxiv.org/abs/hep-ph/0508110} {arXiv:hep-ph/0508110 [hep-ph]}
  \BibitemShut {NoStop}%
\bibitem [{\citenamefont {Munz}(1996)}]{Munz:1996hb}%
  \BibitemOpen
  \bibfield  {author} {\bibinfo {author} {\bibfnamefont {C.~R.}\ \bibnamefont
  {Munz}},\ }\href {\doibase 10.1016/S0375-9474(96)00265-5} {\bibfield
  {journal} {\bibinfo  {journal} {Nucl.Phys.}\ }\textbf {\bibinfo {volume}
  {A609}},\ \bibinfo {pages} {364} (\bibinfo {year} {1996})},\ \Eprint
  {http://arxiv.org/abs/hep-ph/9601206} {arXiv:hep-ph/9601206 [hep-ph]}
  \BibitemShut {NoStop}%
\bibitem [{\citenamefont {Ebert}\ \emph {et~al.}(2003)\citenamefont {Ebert},
  \citenamefont {Faustov},\ and\ \citenamefont {Galkin}}]{Ebert:2003mu}%
  \BibitemOpen
  \bibfield  {author} {\bibinfo {author} {\bibfnamefont {D.}~\bibnamefont
  {Ebert}}, \bibinfo {author} {\bibfnamefont {R.}~\bibnamefont {Faustov}}, \
  and\ \bibinfo {author} {\bibfnamefont {V.}~\bibnamefont {Galkin}},\ }\href
  {\doibase 10.1142/S021773230300971X} {\bibfield  {journal} {\bibinfo
  {journal} {Mod.Phys.Lett.}\ }\textbf {\bibinfo {volume} {A18}},\ \bibinfo
  {pages} {601} (\bibinfo {year} {2003})},\ \Eprint
  {http://arxiv.org/abs/hep-ph/0302044} {arXiv:hep-ph/0302044 [hep-ph]}
  \BibitemShut {NoStop}%
\bibitem [{\citenamefont {Anisovich}\ \emph {et~al.}(2007)\citenamefont
  {Anisovich}, \citenamefont {Dakhno}, \citenamefont {Matveev}, \citenamefont
  {Nikonov},\ and\ \citenamefont {Sarantsev}}]{Anisovich:2005jp}%
  \BibitemOpen
  \bibfield  {author} {\bibinfo {author} {\bibfnamefont {V.}~\bibnamefont
  {Anisovich}}, \bibinfo {author} {\bibfnamefont {L.}~\bibnamefont {Dakhno}},
  \bibinfo {author} {\bibfnamefont {M.}~\bibnamefont {Matveev}}, \bibinfo
  {author} {\bibfnamefont {V.}~\bibnamefont {Nikonov}}, \ and\ \bibinfo
  {author} {\bibfnamefont {A.}~\bibnamefont {Sarantsev}},\ }\href {\doibase
  10.1134/S1063778807010097} {\bibfield  {journal} {\bibinfo  {journal}
  {Phys.Atom.Nucl.}\ }\textbf {\bibinfo {volume} {70}},\ \bibinfo {pages} {63}
  (\bibinfo {year} {2007})},\ \Eprint {http://arxiv.org/abs/hep-ph/0510410}
  {arXiv:hep-ph/0510410 [hep-ph]} \BibitemShut {NoStop}%
\bibitem [{\citenamefont {Wang}(2009)}]{Wang:2009er}%
  \BibitemOpen
  \bibfield  {author} {\bibinfo {author} {\bibfnamefont {G.-L.}\ \bibnamefont
  {Wang}},\ }\href {\doibase 10.1016/j.physletb.2009.03.030} {\bibfield
  {journal} {\bibinfo  {journal} {Phys.Lett.}\ }\textbf {\bibinfo {volume}
  {B674}},\ \bibinfo {pages} {172} (\bibinfo {year} {2009})},\ \Eprint
  {http://arxiv.org/abs/0904.1604} {arXiv:0904.1604 [hep-ph]} \BibitemShut
  {NoStop}%
\bibitem [{\citenamefont {Li}\ and\ \citenamefont {Chao}(2009)}]{Li:2009nr}%
  \BibitemOpen
  \bibfield  {author} {\bibinfo {author} {\bibfnamefont {B.-Q.}\ \bibnamefont
  {Li}}\ and\ \bibinfo {author} {\bibfnamefont {K.-T.}\ \bibnamefont {Chao}},\
  }\href {\doibase 10.1088/0253-6102/52/4/20} {\bibfield  {journal} {\bibinfo
  {journal} {Commun.Theor.Phys.}\ }\textbf {\bibinfo {volume} {52}},\ \bibinfo
  {pages} {653} (\bibinfo {year} {2009})},\ \Eprint
  {http://arxiv.org/abs/0909.1369} {arXiv:0909.1369 [hep-ph]} \BibitemShut
  {NoStop}%
\bibitem [{\citenamefont {Hwang}\ and\ \citenamefont
  {Guo}(2010)}]{Hwang:2010iq}%
  \BibitemOpen
  \bibfield  {author} {\bibinfo {author} {\bibfnamefont {C.-W.}\ \bibnamefont
  {Hwang}}\ and\ \bibinfo {author} {\bibfnamefont {R.-S.}\ \bibnamefont
  {Guo}},\ }\href {\doibase 10.1103/PhysRevD.82.034021} {\bibfield  {journal}
  {\bibinfo  {journal} {Phys.Rev.}\ }\textbf {\bibinfo {volume} {D82}},\
  \bibinfo {pages} {034021} (\bibinfo {year} {2010})},\ \Eprint
  {http://arxiv.org/abs/1005.2811} {arXiv:1005.2811 [hep-ph]} \BibitemShut
  {NoStop}%
\bibitem [{\citenamefont {Abe}\ \emph {et~al.}(1997)\citenamefont {Abe} \emph
  {et~al.}}]{Abe:1997yz}%
  \BibitemOpen
  \bibfield  {author} {\bibinfo {author} {\bibfnamefont {F.}~\bibnamefont
  {Abe}} \emph {et~al.} (\bibinfo {collaboration} {CDF Collaboration}),\ }\href
  {\doibase 10.1103/PhysRevLett.79.578} {\bibfield  {journal} {\bibinfo
  {journal} {Phys.Rev.Lett.}\ }\textbf {\bibinfo {volume} {79}},\ \bibinfo
  {pages} {578} (\bibinfo {year} {1997})}\BibitemShut {NoStop}%
\bibitem [{\citenamefont {Abulencia}\ \emph {et~al.}(2007)\citenamefont
  {Abulencia} \emph {et~al.}}]{Abulencia:2007bra}%
  \BibitemOpen
  \bibfield  {author} {\bibinfo {author} {\bibfnamefont {A.}~\bibnamefont
  {Abulencia}} \emph {et~al.} (\bibinfo {collaboration} {CDF Collaboration}),\
  }\href {\doibase 10.1103/PhysRevLett.98.232001} {\bibfield  {journal}
  {\bibinfo  {journal} {Phys.Rev.Lett.}\ }\textbf {\bibinfo {volume} {98}},\
  \bibinfo {pages} {232001} (\bibinfo {year} {2007})},\ \Eprint
  {http://arxiv.org/abs/hep-ex/0703028} {arXiv:hep-ex/0703028 [HEP-EX]}
  \BibitemShut {NoStop}%
\bibitem [{\citenamefont {Chatrchyan}\ \emph {et~al.}(2012)\citenamefont
  {Chatrchyan} \emph {et~al.}}]{Chatrchyan:2012ub}%
  \BibitemOpen
  \bibfield  {author} {\bibinfo {author} {\bibfnamefont {S.}~\bibnamefont
  {Chatrchyan}} \emph {et~al.} (\bibinfo {collaboration} {CMS Collaboration}),\
  }\href {\doibase 10.1140/epjc/s10052-012-2251-3} {\bibfield  {journal}
  {\bibinfo  {journal} {Eur.Phys.J.}\ }\textbf {\bibinfo {volume} {C72}},\
  \bibinfo {pages} {2251} (\bibinfo {year} {2012})},\ \Eprint
  {http://arxiv.org/abs/1210.0875} {arXiv:1210.0875 [hep-ex]} \BibitemShut
  {NoStop}%
\bibitem [{\citenamefont {Aaij}\ \emph {et~al.}(2012)\citenamefont {Aaij} \emph
  {et~al.}}]{LHCb:2012ac}%
  \BibitemOpen
  \bibfield  {author} {\bibinfo {author} {\bibfnamefont {R.}~\bibnamefont
  {Aaij}} \emph {et~al.} (\bibinfo {collaboration} {LHCb Collaboration}),\
  }\href {\doibase 10.1016/j.physletb.2012.06.077} {\bibfield  {journal}
  {\bibinfo  {journal} {Phys.Lett.}\ }\textbf {\bibinfo {volume} {B714}},\
  \bibinfo {pages} {215} (\bibinfo {year} {2012})},\ \Eprint
  {http://arxiv.org/abs/1202.1080} {arXiv:1202.1080 [hep-ex]} \BibitemShut
  {NoStop}%
\bibitem [{\citenamefont {Aaij}\ \emph {et~al.}(2013)\citenamefont {Aaij} \emph
  {et~al.}}]{Aaij:2013dja}%
  \BibitemOpen
  \bibfield  {author} {\bibinfo {author} {\bibfnamefont {R.}~\bibnamefont
  {Aaij}} \emph {et~al.} (\bibinfo {collaboration} {LHCb Collaboration}),\
  }\href {\doibase 10.1007/JHEP10(2013)115} {\bibfield  {journal} {\bibinfo
  {journal} {JHEP}\ }\textbf {\bibinfo {volume} {1310}},\ \bibinfo {pages}
  {115} (\bibinfo {year} {2013})},\ \Eprint {http://arxiv.org/abs/1307.4285}
  {arXiv:1307.4285} \BibitemShut {NoStop}%
\bibitem [{\citenamefont {{ATLAS
  Collaboration}}(2013)}]{TheATLAScollaboration:2013bja}%
  \BibitemOpen
  \bibfield  {author} {\bibinfo {author} {\bibnamefont {{ATLAS
  Collaboration}}},\ }\href@noop {} {\  (\bibinfo {year} {2013})},\ \bibinfo
  {note} {{ATLAS-CONF-2013-095, ATLAS-COM-CONF-2013-115}}\BibitemShut {NoStop}%
\bibitem [{\citenamefont {Abe}\ \emph {et~al.}(2002)\citenamefont {Abe} \emph
  {et~al.}}]{Abe:2002rb}%
  \BibitemOpen
  \bibfield  {author} {\bibinfo {author} {\bibfnamefont {K.}~\bibnamefont
  {Abe}} \emph {et~al.} (\bibinfo {collaboration} {Belle Collaboration}),\
  }\href {\doibase 10.1103/PhysRevLett.89.142001} {\bibfield  {journal}
  {\bibinfo  {journal} {Phys.Rev.Lett.}\ }\textbf {\bibinfo {volume} {89}},\
  \bibinfo {pages} {142001} (\bibinfo {year} {2002})},\ \Eprint
  {http://arxiv.org/abs/hep-ex/0205104} {arXiv:hep-ex/0205104 [hep-ex]}
  \BibitemShut {NoStop}%
\bibitem [{\citenamefont {Braguta}\ \emph {et~al.}(2009)\citenamefont
  {Braguta}, \citenamefont {Likhoded},\ and\ \citenamefont
  {Luchinsky}}]{Braguta:2008qe}%
  \BibitemOpen
  \bibfield  {author} {\bibinfo {author} {\bibfnamefont {V.V.}~\bibnamefont
  {Braguta}}, \bibinfo {author} {\bibfnamefont {A.K.}~\bibnamefont {Likhoded}}, \
  and\ \bibinfo {author} {\bibfnamefont {A.V.}~\bibnamefont {Luchinsky}},\ }\href
  {\doibase 10.1103/PhysRevD.79.074004} {\bibfield  {journal} {\bibinfo
  {journal} {Phys.Rev.}\ }\textbf {\bibinfo {volume} {D79}},\ \bibinfo {pages}
  {074004} (\bibinfo {year} {2009})},\ \Eprint {http://arxiv.org/abs/0810.3607}
  {arXiv:0810.3607 [hep-ph]} \BibitemShut {NoStop}%
\bibitem [{\citenamefont {Braguta}\ \emph {et~al.}(2008)\citenamefont
  {Braguta}, \citenamefont {Likhoded},\ and\ \citenamefont
  {Luchinsky}}]{Braguta:2008hs}%
  \BibitemOpen
  \bibfield  {author} {\bibinfo {author} {\bibfnamefont {V.V.}~\bibnamefont
  {Braguta}}, \bibinfo {author} {\bibfnamefont {A.K.}~\bibnamefont {Likhoded}}, \
  and\ \bibinfo {author} {\bibfnamefont {A.V.}~\bibnamefont {Luchinsky}},\ }\href
  {\doibase 10.1103/PhysRevD.78.074032} {\bibfield  {journal} {\bibinfo
  {journal} {Phys.Rev.}\ }\textbf {\bibinfo {volume} {D78}},\ \bibinfo {pages}
  {074032} (\bibinfo {year} {2008})},\ \Eprint {http://arxiv.org/abs/0808.2118}
  {arXiv:0808.2118 [hep-ph]} \BibitemShut {NoStop}%
\bibitem [{\citenamefont {Kniehl}\ \emph {et~al.}(2003)\citenamefont {Kniehl},
  \citenamefont {Kramer},\ and\ \citenamefont {Palisoc}}]{Kniehl:2003pc}%
  \BibitemOpen
  \bibfield  {author} {\bibinfo {author} {\bibfnamefont {B.~A.}\ \bibnamefont
  {Kniehl}}, \bibinfo {author} {\bibfnamefont {G.}~\bibnamefont {Kramer}}, \
  and\ \bibinfo {author} {\bibfnamefont {C.~P.}\ \bibnamefont {Palisoc}},\
  }\href {\doibase 10.1103/PhysRevD.68.114002} {\bibfield  {journal} {\bibinfo
  {journal} {Phys.Rev.}\ }\textbf {\bibinfo {volume} {D68}},\ \bibinfo {pages}
  {114002} (\bibinfo {year} {2003})},\ \Eprint
  {http://arxiv.org/abs/hep-ph/0307386} {arXiv:hep-ph/0307386 [hep-ph]}
  \BibitemShut {NoStop}%
\bibitem [{\citenamefont {{CMS Collaboration}}(2013)}]{CMS:2013yha}%
  \BibitemOpen
  \bibfield  {author} {\bibinfo {author} {\bibnamefont {{CMS Collaboration}}},\
  }\href@noop {} {\  (\bibinfo {year} {2013})},\ \bibinfo {note}
  {{CMS-PAS-BPH-13-005}}\BibitemShut {NoStop}%
\end{thebibliography}%
\end{document}